\documentstyle[12pt]{article}
\input psfig.sty
\def\nn{\noindent}
\def\Re{{\cal R \mskip-4mu \lower.1ex \hbox{\it e}\,}}
\def\Im{{\cal I \mskip-5mu \lower.1ex \hbox{\it m}\,}}
\def\ie{{\it i.e.}}
\def\eg{{\it e.g.}}

\def\etal{{\it et al.}}

\def\sub#1{_{\lower.25ex\hbox{$\scriptstyle#1$}}}
\def\tev{\,{\ifmmode\mathrm {TeV}\else TeV\fi}}
\def\gev{\,{\ifmmode\mathrm {GeV}\else GeV\fi}}
\def\mev{\,{\ifmmode\mathrm {MeV}\else MeV\fi}}
\def\to{\rightarrow}

\def\subw{_{\rm w}}
\def\mh{\ifmmode m\sbl H \else $m\sbl H$\fi}
\def\mch{\ifmmode m_{H^\pm} \else $m_{H^\pm}$\fi}
\def\mt{\ifmmode m_t\else $m_t$\fi}
\def\mc{\ifmmode m_c\else $m_c$\fi}
\def\mz{\ifmmode M_Z\else $M_Z$\fi}
\def\mw{\ifmmode M_W\else $M_W$\fi}
\def\mws{\ifmmode M_W^2 \else $M_W^2$\fi}
\def\mhs{\ifmmode m_H^2 \else $m_H^2$\fi}   
\def\mzs{\ifmmode M_Z^2 \else $M_Z^2$\fi}
\def\mts{\ifmmode m_t^2 \else $m_t^2$\fi}
\def\mcs{\ifmmode m_c^2 \else $m_c^2$\fi}
\def\mchs{\ifmmode m_{H^\pm}^2 \else $m_{H^\pm}^2$\fi}
\def\ztwo{\ifmmode Z_2\else $Z_2$\fi}
\def\zone{\ifmmode Z_1\else $Z_1$\fi}
\def\mtwo{\ifmmode M_2\else $M_2$\fi}
\def\mone{\ifmmode M_1\else $M_1$\fi}
\def\tb{\ifmmode \tan\beta \else $\tan\beta$\fi}
\def\xw{\ifmmode x\subw\else $x\subw$\fi}
\def\ch{\ifmmode H^\pm \else $H^\pm$\fi}
\def\lum{\ifmmode {\cal L}\else ${\cal L}$\fi}
\def\inpb{\,{\ifmmode {\mathrm {pb}}^{-1}\else ${\mathrm {pb}}^{-1}$\fi}}
\def\infb{\,{\ifmmode {\mathrm {fb}}^{-1}\else ${\mathrm {fb}}^{-1}$\fi}}
\def\epem{\ifmmode e^+e^-\else $e^+e^-$\fi}
\def\ppb{\ifmmode \bar pp\else $\bar pp$\fi}
\def\bsg{\ifmmode B\to X_s\gamma\else $B\to X_s\gamma$\fi}
\def\bsll{\ifmmode B\to X_s\ell^+\ell^-\else $B\to X_s\ell^+\ell^-$\fi}
\def\bstt{\ifmmode B\to X_s\tau^+\tau^-\else $B\to X_s\tau^+\tau^-$\fi}
\def\lamt{\ifmmode \tilde\lambda\else $\tilde\lambda$\fi}
\def\shat{\ifmmode \hat s\else $\hat s$\fi}
\def\that{\ifmmode \hat t\else $\hat t$\fi}
\def\uhat{\ifmmode \hat u\else $\hat u$\fi}

\newskip\zatskip \zatskip=0pt plus0pt minus0pt
\def\matth{\mathsurround=0pt}

\def\atversim#1#2{\lower0.7ex\vbox{\baselineskip\zatskip\lineskip\zatskip
  \lineskiplimit 0pt\ialign{$\matth#1\hfil##\hfil$\crcr#2\crcr\sim\crcr}}}

\renewcommand{\thefootnote}{\fnsymbol{footnote}}

\begin{document} \begin{titlepage} 
\rightline{\vbox{\halign{&#\hfil\cr
&SLAC-PUB-8071\cr
&March 1999\cr}}}
\begin{center}

{\Large\bf Using Scalars to Probe Theories of Low Scale Quantum Gravity}
\footnote{Work supported by the Department of 
Energy, Contract DE-AC03-76SF00515}
\medskip

\normalsize 
{\large Thomas G. Rizzo } \\
\vskip .3cm
Stanford Linear Accelerator Center \\
Stanford University \\
Stanford CA 94309, USA\\
\vskip .3cm

\end{center}

\begin{abstract} 
Arkani-Hamed, Dimopoulos and Dvali have recently suggested that gravity may 
become strong at energies near 1 TeV which would remove the hierarchy problem. 
Such a scenario can be tested at present and future colliders since the 
exchange of towers of Kaluza-Klein gravitons leads to a set of new 
dimension-8 operators that can play important phenomenological roles. In 
this paper we examine how the production of pairs of scalars at $e^+e^-$, 
$\gamma \gamma$ and hadron colliders can be used to further probe the effects 
of graviton tower exchange. In particular we examine the tree-level 
production of pairs of identical Higgs fields which occurs only at the loop 
level in both the Standard Model and its extension to the Minimal 
Supersymmetric Standard Model. Cross sections for such processes are found to 
be potentially large at the LHC and the next generation of linear colliders. 
For the $\gamma\gamma$ case the role of polarization in improving sensitivity 
to graviton exchange is emphasized.
\end{abstract} 




\renewcommand{\thefootnote}{\arabic{footnote}} \end{titlepage}


\section{Introduction}

Arkani-Hamed, Dimopoulos and Dvali(ADD)~{\cite {nima}} have recently 
proposed a radically interesting solution to the hierarchy problem. ADD 
hypothesize the existence of $n$ additional large spatial dimensions in 
which gravity can live, called `the bulk', whereas all of the fields of the 
Standard Model are constrained to lie on `a wall', which is our 
conventional 4-dimensional world. Gravity only appears to be weak in our 
ordinary 4-dimensional space-time since we merely observe it's action on the 
wall. It has recently been shown~{\cite {nima}} that a 
scenario of this type may emerge in 
string models where the effective Planck scale in the bulk is identified 
with the string scale. In such a theory the hierarchy can be 
removed by postulating that the string or effective Planck scale in 
the bulk, $M_s$, is not far above the weak scale, \eg, a few TeV. Gauss' Law 
then provides a link between the values of $M_s$, the conventional 
Planck scale $M_{pl}$, and the size of the compactified extra dimensions, $R$, 
\begin{equation}
M_{pl}^2 \sim R^nM_s^{n+2}\,,
\end{equation}
where the constant of proportionality depends not only on the value of $n$ 
but upon the geometry of the compactified dimensions. Interestingly, if $M_s$ 
is near a TeV then $R\sim 10^{30/n-19}$ meters; for separations between two 
masses less than $R$ the gravitational 
force law becomes $1/r^{2+n}$. For $n=1$, 
$R\sim 10^{11}$ meters and is thus obviously excluded, but, for $n=2$ one 
obtains $R \sim 1$~mm, which is at the edge of the sensitivity for existing 
experiments{\cite {test}}. For $2<n \leq 7$, where 7 is the maximum value 
of $n$ 
being suggested by M-theory, the value of $R$ is further reduced and thus we 
may conclude that the range $2\leq n \leq 7$ is of phenomenological interest. 
Astrophysical arguments{\cite {astro}} suggest that $M_s>50$ TeV for $n=2$, 
but allow $M_s\sim 1$ TeV for $n>2$. 

The Feynman rules for this theory are obtained by considering 
a linearized theory of gravity 
in the bulk, decomposing it into the more familiar 4-dimensional states and 
recalling the existence of Kaluza-Klein towers for each of the conventionally 
massless fields. The entire set of fields in the K-K tower couples in an 
identical fashion to the particles of the SM. By considering the forms of the 
$4+n$  
symmetric conserved stress-energy tensor for the various SM fields and by 
remembering that such fields live only on the wall one may derive all of the 
necessary couplings. An important result of 
these considerations is that only the massive spin-2 K-K towers (which couple 
to the 4-dimensional stress-energy tensor, $T^{\mu\nu}$) and spin-0 K-K 
towers (which couple proportional to the trace of $T^{\mu\nu}$) are of 
phenomenological relevance as all the spin-1 fields can be shown to decouple 
from the particles of the SM. For processes that involve massless fields at 
at least one vertex the contributions of the spin-0 fields can also be ignored.

The details of the phenomenology of the ADD model has begun to be explored in 
a series of recent papers~{\cite {pheno}}. 
Given the Feynman rules as developed by Guidice, Rattazzi and Wells and by 
Han, Lykken and Zhang{\cite {pheno}}, it appears that the 
ADD scenario has two basic classes of collider tests. In the first class, 
type-$i$, 
a K-K tower of gravitons can be emitted during a decay or scattering process 
leading to a final state with missing energy. The rate for such processes is 
strongly dependent on the number of extra dimensions as well as the exact 
value of $M_s$. In the second class, type-$ii$, 
which we consider here, the exchange of a K-K graviton tower between SM or 
MSSM fields can lead to almost 
$n$-independent modifications to conventional cross sections and 
distributions or they can possibly lead to new interactions. The exchange of 
the graviton K-K tower leads to a set of effective color and flavor singlet 
contact interaction operator of dimension-eight with the scale set by the 
parameter $M_s$. The universal overall order one coefficient of these 
operators, $\lambda$, is unknown but its value is conventionally 
set to $\pm 1$. Given the kinematic structure of these operators the 
modifications in the relevant 
cross sections and distributions can be directly calculated. 

Until now the phenomenological analyses have concentrated{\cite {pheno}} on 
the collider processes $f_1\bar f_2\to f_3\bar f_4,VV$ and 
$\gamma\gamma,gg\to f\bar f$, where the $f_i$'s are 
fermions and $V$'s are vector bosons, since these are the traditional 
components of the SM. However in SUSY models, in particular in the case of the 
Minimal Supersymmetric Standard Model(MSSM), scalar fields exist on the same 
footing as do fermions and vectors. Therefore in this paper we turn to a 
complementary examination of the influence of the exchange of a tower of K-K 
gravitons on the processes $e^+e^-,\gamma\gamma \to S\bar S$ where $S$ is a 
real or complex scalar. From the nature of these new operators the application 
to the processes $q\bar q,gg \to S\bar S$ becomes immediately apparent. 
While we do not expect the existence of such K-K 
exchanges to be discovered in these channels the processes we consider do 
offer new ways to probe 
this phenomena and possibly confirming information through the universality 
expected in the couplings of low scale quantum gravity. As we will see, many 
of the modifications of cross sections and asymmetries due to K-K tower 
exchange observed earlier in the case of fermion final states will have their 
parallel in the case of scalar final states. Of particular 
interest, as we will see below, is the tree level production of identical 
pairs of neutral Higgs 
bosons via K-K tower exchange in $e^+e^-$, $\gamma \gamma$ as well as hadronic 
collisions with rather unique kinematic properties. In the SM or MSSM these 
processes can only occur at the loop level with rather small cross sections.

\section{$e^+e^- \to$ Scalar Pairs}

The production of pairs of scalars is a basic elementary $e^+e^-$ process in 
the MSSM. For the reaction $e^-(k_1)e^+(k_2)\to S(p_1)\bar S(p_2)$ [with the 
$k_i$($p_i$) incoming(outgoing)], a K-K tower of gravitons contributes an 
additional amplitude of the form 
\begin{equation}
{\cal M}={2\lambda K\over M_s^4}(t-u)(p_1-p_2)_\mu\bar e(k_2)
\gamma^\mu e(k_1)\,,
\end{equation}
with $u,t$ being the usual Mandelstam kinematic variables. 
Thus, including the contributions from $\gamma$, $Z$ and K-K graviton tower 
exchanges, the full 
differential cross section for $e^+e^-\to S\bar S$ is given compactly by 
\begin{eqnarray}
{d\sigma \over {dz}}&=& N_c{\pi \alpha^2 \over {4s}}\beta^3(1-z^2)\bigg[
\sum_{ij} c_ic_j(v_iv_j+a_ia_j)P_iP_j \nonumber \\
&-& 2 C\beta z \sum_{i} c_iv_iP_i + C^2 (\beta z)^2\bigg]\,,
\end{eqnarray}
where $z=\cos \theta$, $s$ is the square of the collider center of mass 
energy, $N_c$ is the color factor for the scalar in the final 
state, $v_i,a_i$ are the electron's vector and axial vector 
couplings to the gauge boson $i(=\gamma,Z$), $c_i$ are the corresponding 
couplings for the scalar $S$, $P_i=s/(s-M_i^2)$ are propagator factors for 
the gauge bosons of mass $M_i$, $C=\lambda Ks^2/2\pi \alpha M_s^4$ and 
$\beta^2=1-4m_S^2/s$, 
with $m_S$ being the scalar mass. Our couplings are normalized such that 
$v_\gamma=-1$ and $c_\gamma=Q$, the scalar's electric charge. For 
$K=1(\pi/2)$ we 
recover the normalization convention employed by Hewett(Guidice, Rattazzi and 
Wells){\cite {pheno}}; we will take $K=1$ in the numerical analysis that 
follows but keep the factor in our analytical expressions. 
We recall from the Hewett analysis that $\lambda$ is 
a parameter of order unity whose sign is undetermined and that, given the 
scaling relationship between $\lambda$ and $M_s$, experiments in the 
case of processes 
of type-$ii$ actually probe only the combination $M_s/|K\lambda|^{1/4}$. 
For simplicity in what follows we will numerically set $|\lambda|=1$ and 
employ $K=1$ but we caution the reader about this 
technicality and quote our sensitivity to $M_s$ for $\lambda=\pm 1$. 

Note that in the SM or MSSM the well-known scalar pair production cross 
section angular 
distribution behaves as $\sim \sin^2 \theta$ as expected. However, the exchange 
of a K-K-tower of gravitons can alter this distribution in several 
statistically significant ways provided $s/M_s^2$ is not too small. 
First, both the interference and pure graviton terms lead to higher powers 
of $z$ in the angular distribution as was first 
observed in the case of fermion pair 
final states by Hewett{\cite {pheno}}. Second, the interference term leads to 
a forward-backward asymmetry, $A_{FB}$, in the angular distribution--something 
not generally expected for scalar pair production. Note that this interference 
term, being odd in $z$, vanishes upon integration and makes no contribution 
to the total cross section, as was also 
the case for fermion pair production. This 
implies that the total cross section can only increase with respect to the 
expectations of the SM when graviton exchanges are present. 

\vspace*{-0.5cm}
\nn
\begin{figure}[htbp]
\centerline{
\psfig{figure=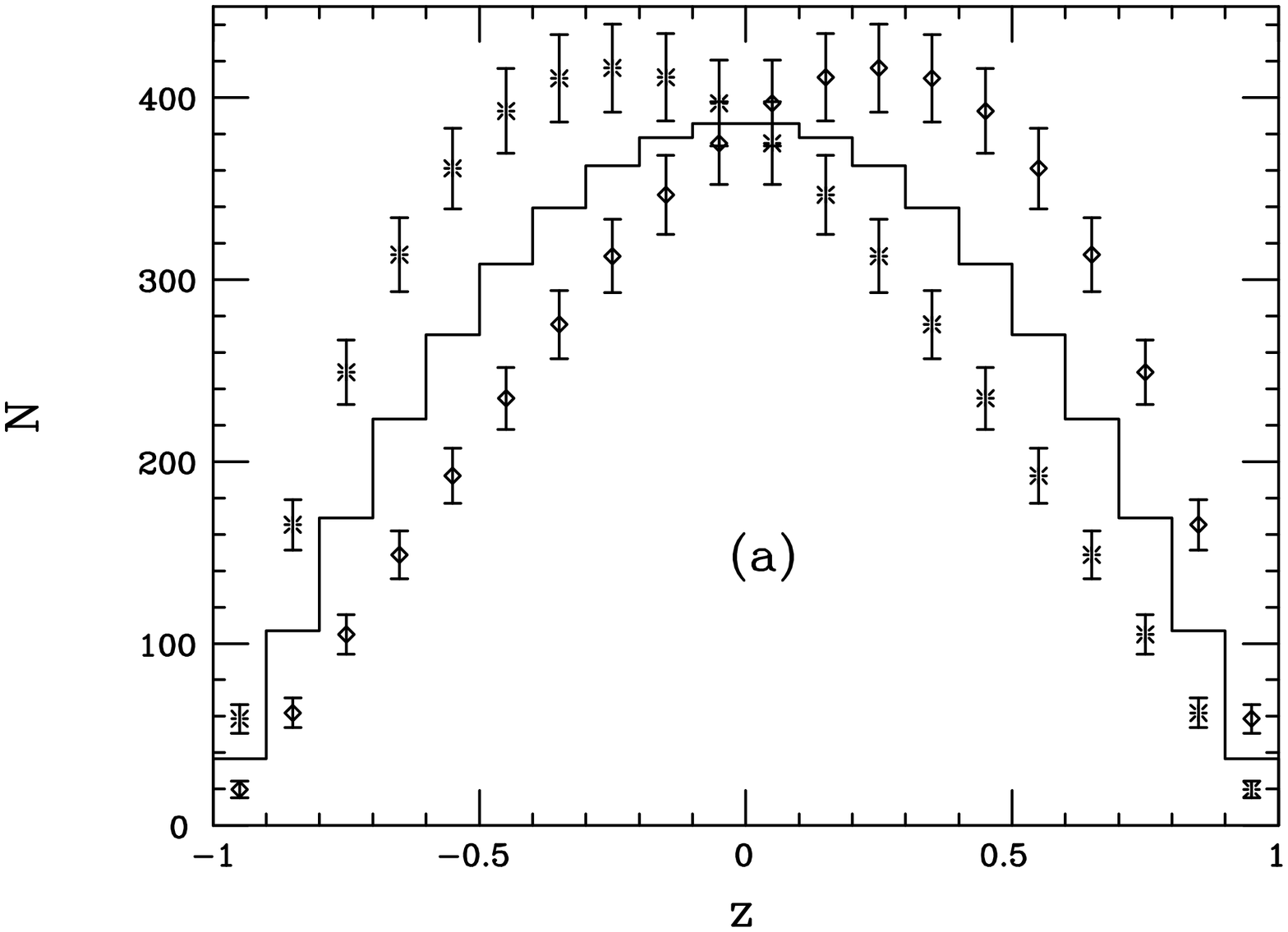,height=10.5cm,width=14cm,angle=-90}}
\vspace*{-10mm}
\centerline{
\psfig{figure=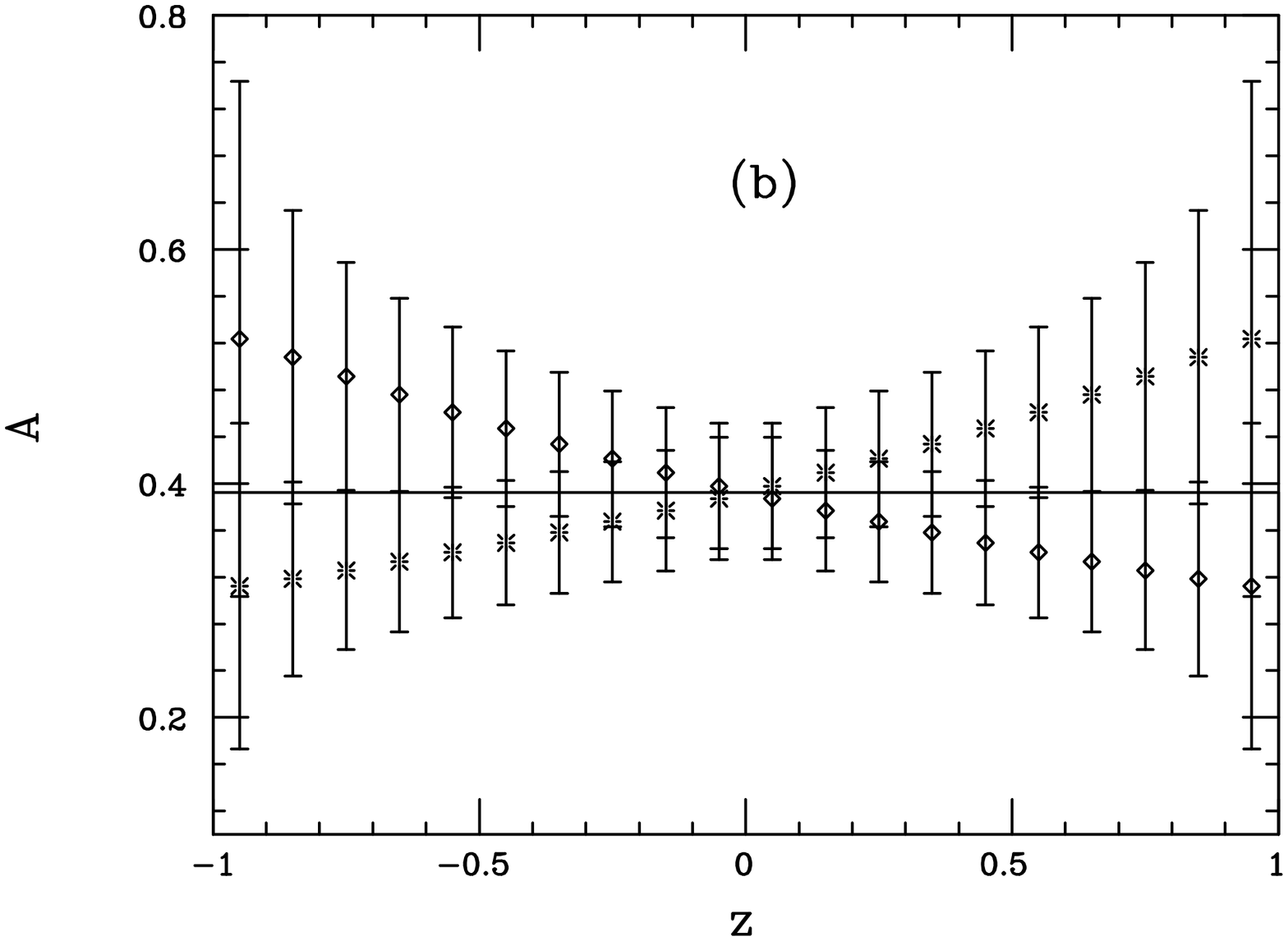,height=10.5cm,width=14cm,angle=-90}}
\vspace*{-0.9cm}
\caption{(a)Cross section and (b)$A_{LR}$ for $\tilde t$ production at a 500 
GeV $e^+e^-$ collider assuming $m_{\tilde t}=150$ GeV and an integrated 
luminosity of 100 $fb^{-1}$. The stop is assumed to be maximally mixed, \ie, 
$\theta_{\tilde t}=45^o$. The histogram(solid line) shows the MSSM 
expectations while the two sets of data points include the contributions 
from a K-K tower of gravitons with 
$M_s=1.5$ TeV and reflect the anticipated measurement errors.}
\label{fig1}
\end{figure}
\vspace*{0.4mm}

With at least single $e^-$ beam polarization one can define a $z$-dependent 
Left-Right asymmetry, $A_{LR}(z)$, associated with scalar pair production given 
by
\begin{equation}
A_{LR}(z)={(1-z^2)\bigg[\sum_{ij} c_ic_j(v_ia_j+a_iv_j)P_iP_j 
-2C\beta z \sum_{i} c_ia_iP_i\bigg]\over {(1-z^2)
\bigg[\sum_{ij} c_ic_j(v_iv_j+a_ia_j)P_iP_j 
-2C\beta z \sum_{i} c_iv_iP_i + C^2 (\beta z)^2\bigg]}}\,, 
\end{equation}
where the expression in the denominator is essentially that for the 
differential cross section above. 
Note that in the case of the SM, MSSM or in any model with an additional $Z'$ 
exchange, the $z$ dependence completely cancels and $A_{LR}$ becomes a 
constant for a fixed value of $\sqrt s$. A third effect of the exchange of 
K-K-towers is 
thus to give $A_{LR}$ a finite (approximately odd) $z$ dependence which may 
be observable 
provided the ratio $s/M_s^2$ is not too small. Combined with the first two 
effects described above we see that the exchange of a K-K-tower of gravitons 
leads to rather unique signatures in the case of the production of scalar 
pairs. The angular-averaged value of $A_{LR}(z)$ can also be obtained from 
this expression by separately integrating the numerator and denominator  of 
Eq.4 over $z$. In a similar fashion we note that the odd terms in $z$ 
contained in $A_{LR}(z)$ can be directly probed by forming an integrated 
Left-Right Forward-Backward asymmetry, $A^{LR}_{FB}$, by analogy with the 
more conventional integrated Forward-Backward asymmetry, $A_{FB}$. We remind 
the reader that both of these quantities are expected to be zero in the MSSM 
but will now differ from zero due to K-K tower exchange by terms of order 
$s^2/M_s^4$. 

\vspace*{-0.5cm}
\nn
\begin{figure}[htbp]
\centerline{
\psfig{figure=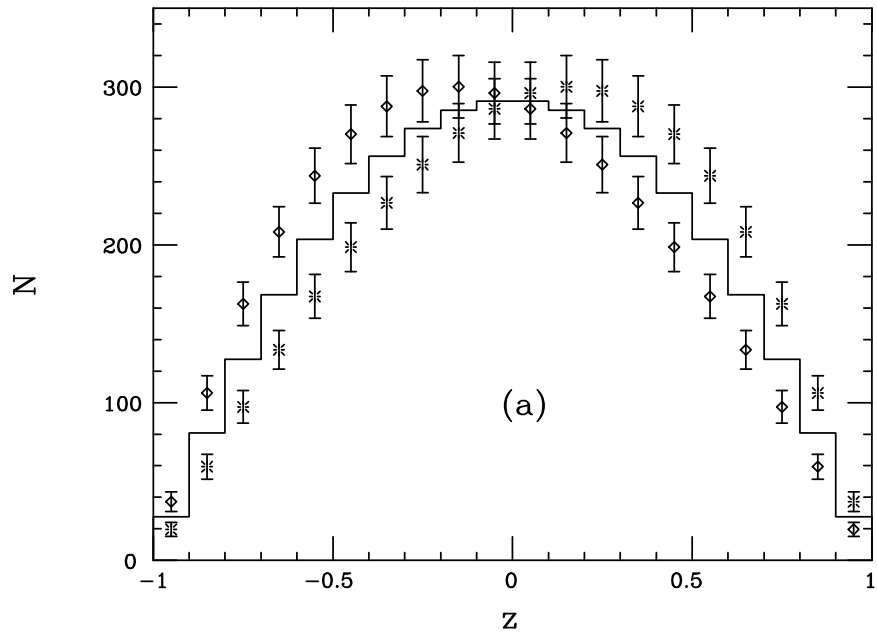,height=10.5cm,width=14cm,angle=-90}}
\vspace*{-10mm}
\centerline{
\psfig{figure=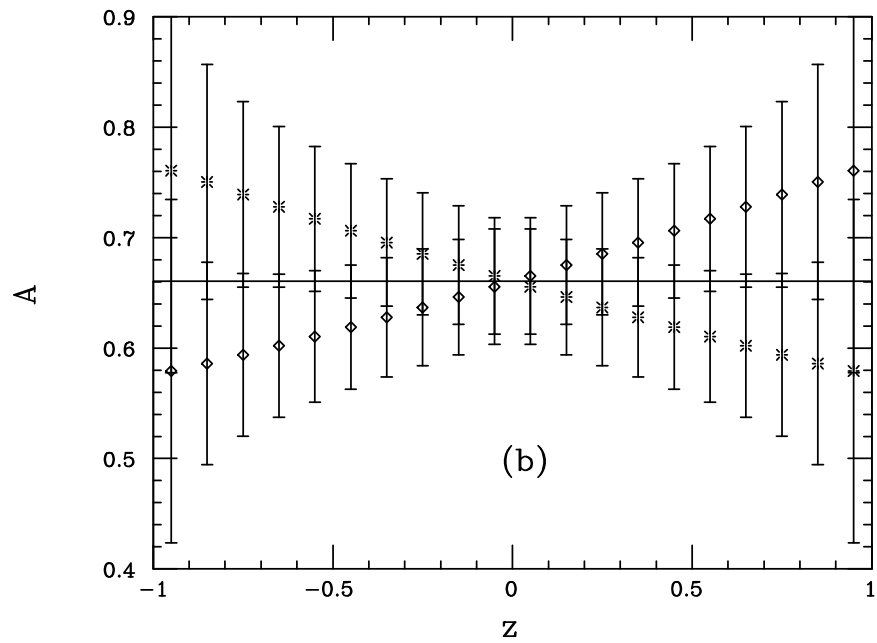,height=10.5cm,width=14cm,angle=-90}}
\vspace*{-0.9cm}
\caption{Same as the previous figure but now for either a $\tilde \mu_L$ or a 
charged Higgs boson, $H^\pm$. Note that the slope of $A_{LR}$ in this case is 
opposite to that for $\tilde \mu_L$ or $H^\pm$.}
\label{fig2}
\end{figure}
\vspace*{0.4mm}

At this point it is informative to go through a few typical examples. In the 
MSSM the scalars we have available to study are the SUSY partners of the 
quarks and leptons as well as the five physical Higgs fields. For purposes 
of demonstration, we consider 
the pair production of these fields at a 500 GeV $e^+e^-$ collider. In this 
case, except for the possibility of highly mixed stop squark($\tilde t$), 
we expect that the squarks will 
most likely be too massive to be pair produced. Thus we will consider the 
pair production of scalars for the 
following four cases: ($i$) A maximally mixed light $\tilde t_1$, \ie, where 
the stop mixing angle between $\tilde t_L$ and $\tilde t_R$ which determines 
the $Z\tilde t_1 \tilde t_1^*$  is given by $\theta_{\tilde t}=45^o$,  
($ii$) a charged 
Higgs or $\tilde \mu_L$ (they have identical electroweak couplings), ($iii$) 
a $\tilde \mu_R$ and ($iv$) the production of a pair of identical neutral 
Higgs bosons. 
This last possibility is particularly interesting since it cannot occur at the 
tree level in the MSSM. Instead of $\tilde \mu_{L,R}$, one could just as well 
examine selectron pair production 
but in that case the additional $t$-channel graphs make the 
analysis and the influence of the K-K tower contributions less transparent.

\vspace*{-0.5cm}
\nn
\begin{figure}[htbp]
\centerline{
\psfig{figure=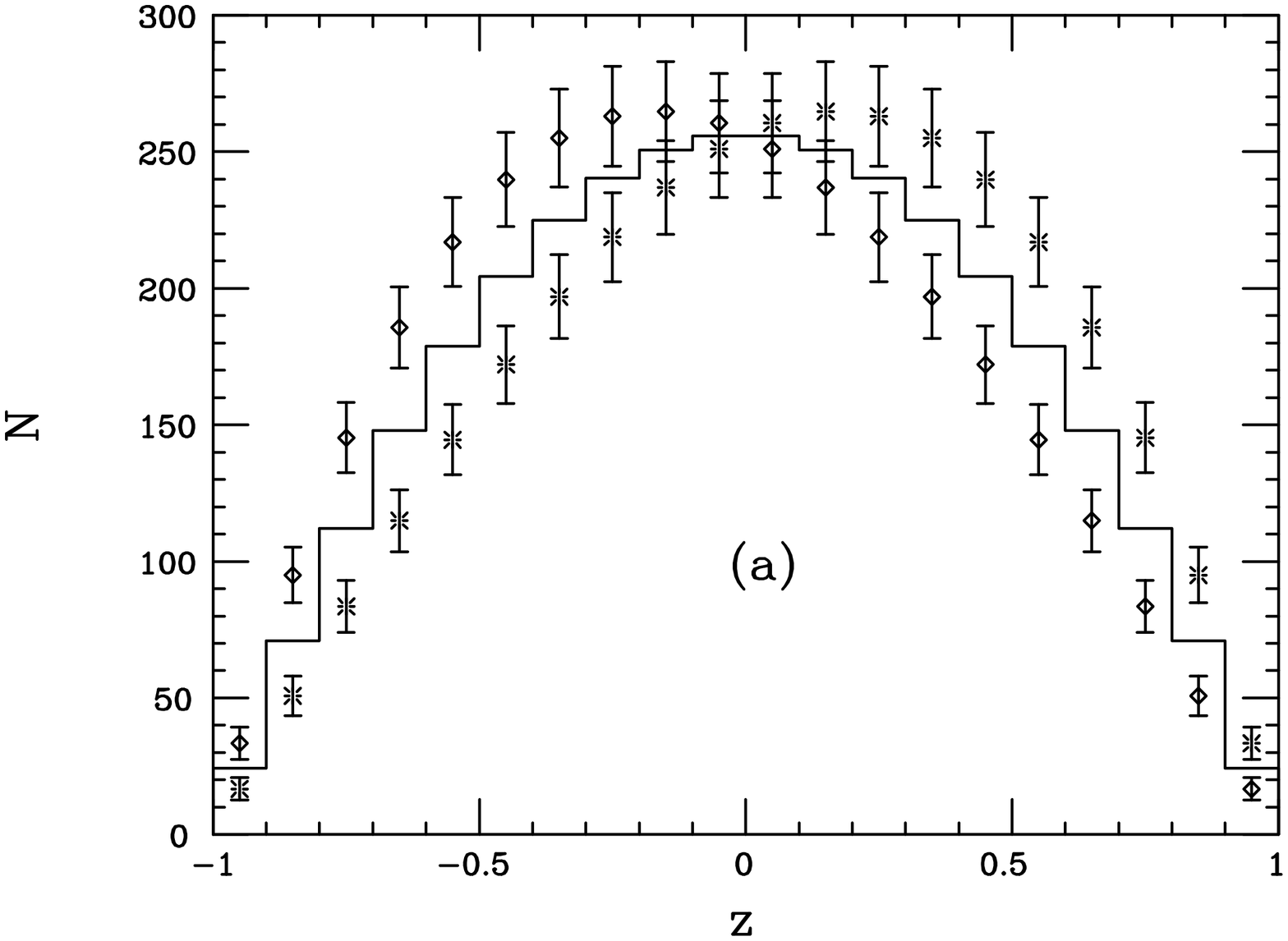,height=10.5cm,width=14cm,angle=-90}}
\vspace*{-10mm}
\centerline{
\psfig{figure=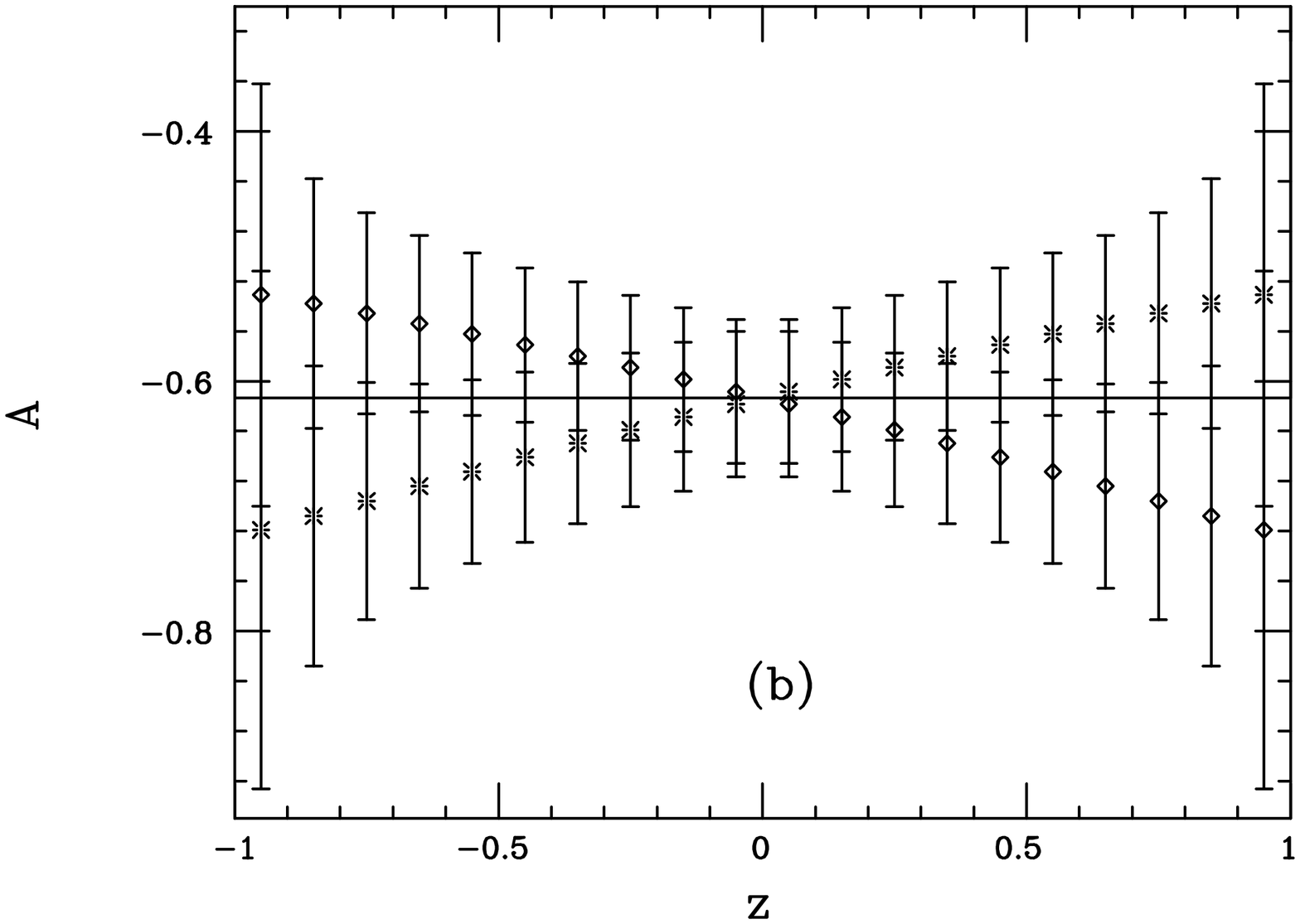,height=10.5cm,width=14cm,angle=-90}}
\vspace*{-0.9cm}
\caption{Same as the previous figure but now for a $\tilde \mu_R$.}
\label{fig3}
\end{figure}
\vspace*{0.4mm}

Fig.\ref{fig1} shows the angular distribution for the production of 150 GeV 
maximally mixed stop squark pairs at a 500 GeV $e^+e^-$ collider assuming an 
integrated luminosity of 100 $fb^{-1}$. In the absence of contributions from 
low scale quantum gravity the distribution is forward-backward symmetric, 
$\sim \sin^2 \theta$. When the K-K tower exchange contribution is turned on it 
leads to a skewing of the distribution to either the forward or backward 
direction depending upon the sign of $\lambda$. Overall there is not a large 
change in the total cross section, $\sim 2\%$ in the example shown in the 
figure. However, in this case the value of $A_{FB}$ differs from zero by many 
$\sigma$ as one might expect from a simple visual inspection of the angular 
distribution; for this integrated luminosity 
we obtain $A_{FB}=\pm 0.210\pm 0.013$, where the overall sign is 
directly correlated with the sign of $\lambda$. 
Also one sees that $A_{LR}$ picks up a $z$ 
dependence as promised, tilting the distribution into either the forward or 
backwards direction. While the angular averaged value of $A_{LR}$ differs from 
the MSSM expectation by less than $1\%$ since it is almost purely odd in $z$, 
the integrated Left-Right 
Forward-Backward asymmetry, $A^{LR}_{FB}$, differs from zero by 
approximately $\sim 3\sigma$ with this assumed integrated luminosity; we 
obtain $A^{LR}_{FB}=\pm 0.043\pm 0.015$ with the sign the same as that of 
$A_{FB}$. Combining all of the observables and 
assuming $100\%$ efficiencies, the lack of observation of such effects would 
place a $95\%$ CL lower bound of $\simeq 4.65$ TeV on $M_s$, which is 
comparable to the {\it discovery} reaches obtained using the more conventional 
fermionic channels{\cite {pheno}} 
at this center of mass energy and assumed integrated luminosity.

\vspace*{-0.5cm}
\nn
\begin{figure}[htbp]
\centerline{
\psfig{figure=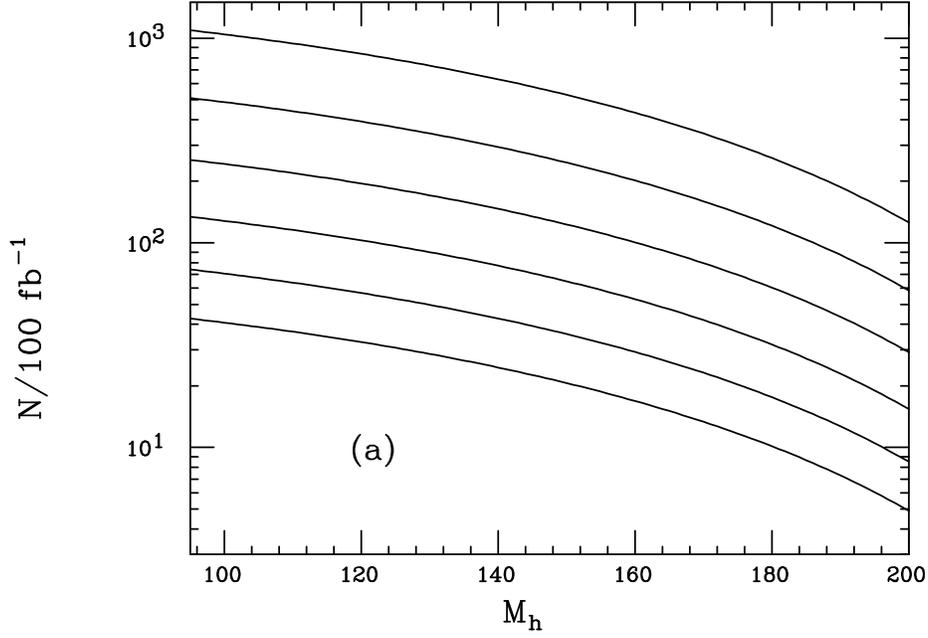,height=10.5cm,width=14cm,angle=-90}}
\vspace*{-10mm}
\centerline{
\psfig{figure=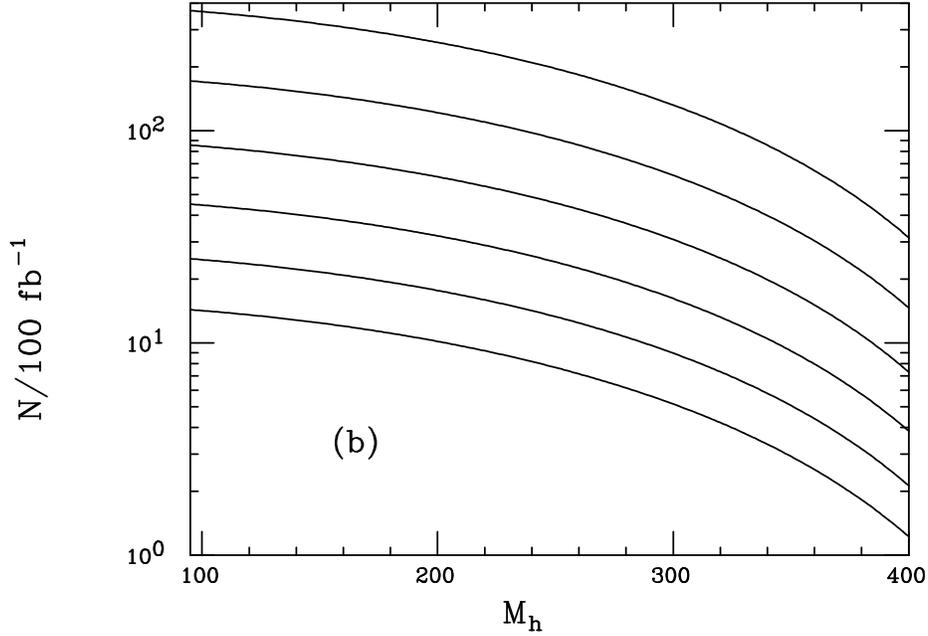,height=10.5cm,width=14cm,angle=-90}}
\vspace*{-0.9cm}
\caption{Tree level production rate for Higgs boson pairs due to graviton tower 
exchange at a (a)500 GeV or a (b)1 TeV $e^+e^-$ 
collider as a function of the Higgs mass scaled to an integrated luminosity 
of 100 $fb^{-1}$. From top to bottom in (a)[(b)] the curves correspond to the 
choice $M_s=1[2]$ TeV increasing in steps of 0.1[0.2] TeV.}
\label{fig4}
\end{figure}
\vspace*{0.4mm}

Figs.\ref{fig2} and \ref{fig3} show that the case of stop pair production is 
not in any way special. Although the overall effect of the exchange of the K-K 
tower of gravitons is now somewhat softer in the case of sleptons and charged 
Higgs and less statistics is available due to the smaller cross sections, it 
is clearly present in both of 
these figures. In both these cases, the graviton tower exchange increases the 
total cross section by less than $1\%$ and shifts the integrated value of 
$A_{LR}$ by a comparable amount. In the $\tilde \mu_L(H^-)$ case we find, 
however, that  $A_{FB}=\pm 0.127\pm 0.016$ and 
$A^{LR}_{FB}=\pm 0.048\pm 0.018$ with common signs. In the $\tilde \mu_R$ 
example, we find instead that  $A_{FB}=\pm 0.137\pm 0.017$ and 
$A^{LR}_{FB}=\mp 0.047\pm 0.019$ with the signs {\it anti-correlated}. The 
bounds that one could potentially obtain on $M_s$ in either one of these two 
examples would be somewhat smaller than that obtainable from $\tilde t$ pair 
production; for both cases we obtain at $95\%$ CL the bound of 3.40 TeV 
assuming 100 $fb^{-1}$ of integrated luminosity and perfect efficiencies. 

\vspace*{-0.5cm}
\nn
\begin{figure}[htbp]
\centerline{
\psfig{figure=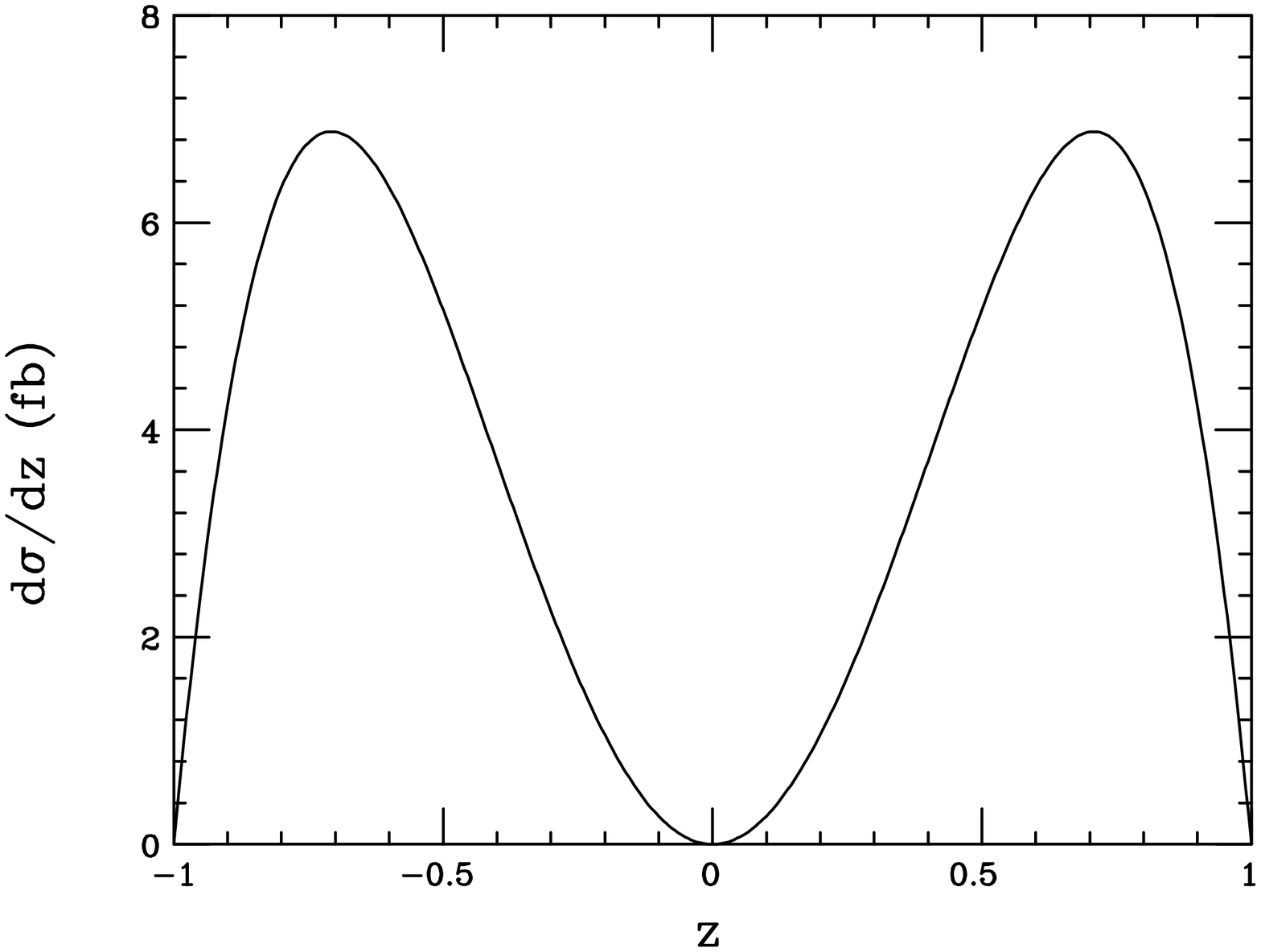,height=14cm,width=17cm,angle=-90}}
\vspace*{-1cm}
\caption[*]{The tree level unpolarized differential 
cross section for $e^+e^-\to 2h^0$ at a 500 GeV $e^+e^-$ linear collider due 
to the exchange of a Kaluza-Klein tower of gravitons 
assuming $m_h=130$ GeV and $M_s=1$ TeV. Note the canonical shape arising from 
the nature of the spin-2 exchange.}
\label{fig3p}
\end{figure}
\vspace*{0.4mm}

One interesting consequence of a K-K graviton tower exchange is the existence 
of new operators that can lead to unusual tree-level processes such as 
reaction $e^+e^-\to 2h^0$, where $h^0$ can be the SM Higgs, the light or 
heavy CP-even Higgs of the MSSM, $h,H$, or the corresponding CP-odd field $A$. 
As we will comment upon in more detail below, such processes can indeed occur 
in the SM or MSSM but only at loop 
level{\cite {loop1}}. Unlike the MSSM case, however, both fields in the final 
state {\it must} be identical, \eg, the $hH$ or $hA$ final states are not 
accessible. The cross section can easily be obtained from the 
expressions above by turning off the photon and $Z$ contributions and dividing 
by a factor of 2 since identical particles are being produced in the final 
state. We find, allowing for polarization of either beam, 
\begin{equation}
{d\sigma \over {dz}}= {\lambda^2K^2s^3\over {32\pi M_s^8}}\beta^5z^2
(1-z^2)(1-\lambda_{e^+}\lambda_{e^-})\,,
\end{equation}
where the $\lambda_{e^\pm}$ are the positron and electron helicities, 
respectively. The unpolarized cross section is obtained by averaging over all 
four helicity combinations; note that both the $e_L^-e_R^+\to 2h^0$ and the 
$e_R^-e_L^+\to 2h^0$ cross sections due to K-K graviton exchange are predicted 
to be identical. 

Another source of background to the $hh$ signal arises from $Zh$ channel, 
which is likely to be used to discover a Higgs boson at a linear collider. 
Our analysis assumes that the Higgs will already have been discovered and its 
mass well determined long before one begins to look for the $hh$ final state 
arising from extra dimensions. Knowing the Higgs mass and the $\sim 1-z^2$ 
angular distribution associated withe the $Zh$ channel, 
it should be possible to eliminate this source of 
background completely especially if the Higgs is more massive than 95 GeV,  
which now appears to be the case from searches at LEP II.

In this discussion we have assumed that the Higgs boson is not in any way 
special due to its role as the source of electroweak symmetry breaking. It 
is possible to imagine that the $hh$-graviton coupling is non-canonical in 
some way which may alter the predictions above in detail but not in any 
qualitative manner. 

In a number of ways the K-K graviton-induced cross section for Higgs pair 
production can be easily distinguished from the loop-induced SM or MSSM 
contribution{\cite {loop1}}. First, for light Higgs, both the SM and 
MSSM cross sections are 
quite small, of order 0.1-0.2 fb for $\sqrt s=500$ GeV, and have a rough 
$\sim \sin^2 \theta$ angular distribution. In the K-K case, as shown in 
Fig.\ref{fig4}, this small of a 
cross section is only obtained when $\sqrt s/M_s < 1/4$. The shape of the 
angular distribution in the case of graviton exchange is also quite 
distinctive,  
as shown in Fig.\ref{fig3p}, owing to the nature of the spin-2 exchange. 
Secondly, while graviton exchange leads to identical cross sections for both 
$e_L^-e_R^+\to 2h^0$ and $e_R^-e_L^+\to 2h^0$ processes, these are found to 
differ by a factor of 2 in both the SM and MSSM cases. Lastly, as stated 
above, the graviton induced cross section is the same for $hh$, $HH$ and $AA$ 
final states with the same mass which will not necessarily be 
the case in the MSSM.

\section{$\gamma \gamma \to$ Scalar Pairs}

$\gamma \gamma$ collisions offer a unique and distinct window on the 
possibility of new physics in a particularly clean environment. At tree 
level the cross section for particle pair production depends only upon QED-like 
couplings and is thus independent of many other factors such as weak isospin 
and various mixing parameters. Unlike $e^+e^-$ collisions, however, Bose 
symmetry forbids the existence of non-zero values for either $A_{FB}$ or 
$A^{LR}_{FB}$, which were powerful weapons in probing for K-K graviton tower 
exchanges. In the case of $\gamma\gamma$ collisions our remaining tools are 
the angular distributions of the produced scalar pairs and their sensitivity to 
the polarization of the initial state photons.

Polarized $\gamma\gamma$ collisions may be possible at future $e^+e^-$ 
colliders through the use of Compton backscattering of polarized low energy 
laser beams off of polarized high energy electrons{\cite {telnov}}. The 
backscattered photon distribution, $f_\gamma(x=E_\gamma/E_e)$, is far from 
being monoenergetic and is cut off above $x_{max}\simeq 0.83$ implying that 
the colliding photons are significantly softer than the parent lepton beam 
energy. As we will see, this cutoff at 
large $x$, $x_{max}$, implies that the $\gamma \gamma$ center of mass energy 
never exceeds $\simeq 0.83$ of the parent collider and this will result in a 
significantly degraded $M_s$ sensitivity. In 
addition, the shape of the function $f_\gamma$ is somewhat sensitive to the 
polarization state of both the initial laser ($P_l$) and electron ($P_e$) 
whose values fix the specific distribution. While it is anticipated that the 
initial laser polarization will be near $100\%$, \ie, $|P_l|=1$, the electron 
beam polarization is expected to be be near $90\%$, \ie, $|P_e|=0.9$. We will 
assume these values in the analysis that follows. With two photon `beams' 
and the choices $P_l=\pm 1$ and $P_e=\pm 0.9$ to be made for each beam it 
would appear that 16 distinct polarization-dependent cross sections need to be 
examined. However, due to the exchange symmetry between the two photons and the 
fact that a simultaneous flip in the signs of all the polarizations leaves the 
product of the fluxes and the cross sections invariant, we find that there 
only six physically 
distinct polarization combinations. In what follows we will label these 
possibilities by the corresponding signs of the electron and laser 
polarizations as $(P_{e1},P_{l1},P_{e2},P_{l2})$, For example, the 
configuration $(-++-)$ corresponds to $P_{e1}=-0.9$, $P_{l1}=+1$, $P_{e2}=0.9$ 
and $P_{l2}=-1$. 

Clearly some of these polarization 
combinations will be more sensitive to the effects of K-K towers of gravitons 
than will others so our analysis can be used to pick out those particular 
cases.

\vspace*{-0.5cm}
\nn
\begin{figure}[htbp]
\centerline{
\psfig{figure=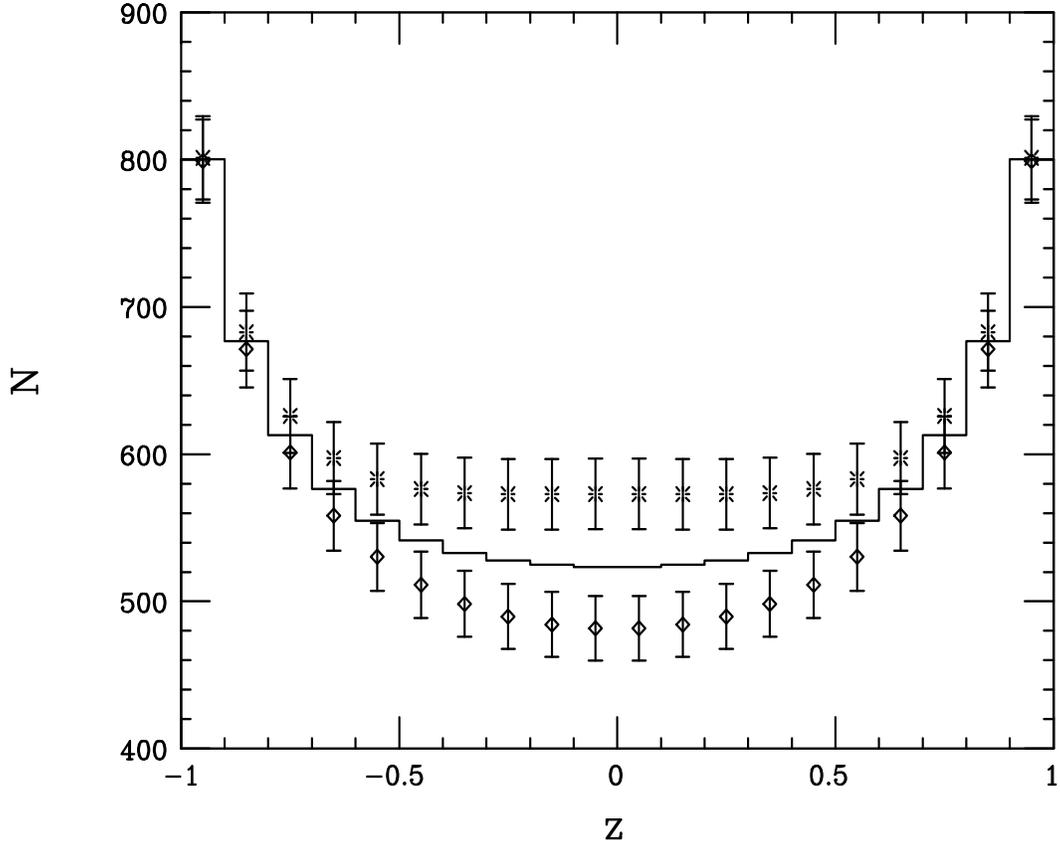,height=14cm,width=17cm,angle=-90}}
\vspace*{-1cm}
\caption[*]{Binned angular distribution for stop pairs produced in unpolarized 
$\gamma \gamma$ collisions assuming an $e^+e^-$ center of mass energy of 
1 TeV for a stop mass of 250 GeV and an integrated luminosity of 200 $fb^{-1}$. 
The histogram shows the result in the MSSM while the two sets of `data' are 
the predictions including the exchange of a tower of K-K gravitons with 
$\lambda=\pm 1$ and $M_s=2.5$ TeV. The errors shown are statistical only.}
\label{fig5}
\end{figure}
\vspace*{0.4mm}

For the reaction $\gamma(k_1)\gamma(k_2)\to S(p_1)\bar S(p_2)$ [with the 
$k_i$($p_i$) incoming(outgoing)], the K-K tower of gravitons contributes an 
additional amplitude of the form 
\begin{eqnarray}
{\cal M} & = & {8\lambda K\over {M_s^4}} \bigg[ 2k_1\cdot k_2 \left \{ p_1\cdot 
\epsilon_1 p_2\cdot \epsilon_2 +p_1\cdot \epsilon_2 p_2\cdot \epsilon_1 
\right \}+2\epsilon_1 \cdot \epsilon_2  \nonumber \\
&  & \left \{ k_1\cdot p_1 k_2\cdot p_2 +k_1\cdot p_2 k_2\cdot p_1 
-k_1\cdot k_2 p_1\cdot p_2\right \}\bigg] \,,
\end{eqnarray}
following our earlier notation and with the $\epsilon_i$ being the 
polarization vectors of the incoming photons. 
The full subprocess cross section for polarized $\gamma \gamma \to S\bar S$ 
reaction can now be written as

\begin{eqnarray}
{d\hat \sigma \over {dz}}&=& N_c{\pi \alpha^2 \over {2\hat s}}\beta 
\bigg[\{(Y-Q^2)^2+[Q^2(1-2X)+Y]^2\}\nonumber \\
&-& 2\xi_1\xi_2(Y-Q^2)[Q^2(1-2X)+Y]\bigg]\,,
\end{eqnarray}
where the $\xi_i$ are the photon helicities{\cite {chak}} and 
\begin{eqnarray}
 X &=& {\beta^2(1-z^2)\over {1-\beta^2z^2}}\nonumber \\
 Y &=& {\lambda K \hat s^2\beta^2 (1-z^2)\over {4\pi \alpha M_s^4}}\,,
\end{eqnarray}
with $z=\cos \theta$, where $\theta$ being the partonic center of mass 
scattering angle as above, $\hat s$ is the square of the sub-process center 
of mass energy, $\beta$ is defined as above but now with the replacement, 
$s\to \hat s$,  and $Q$ is the scalar's electric charge. Note that the 
contribution due to graviton tower exchange is largest when $z=0$, \ie, at 
scattering angles of $90^o$. To 
obtain the corresponding unpolarized cross section we simply average over the 
values of the photon helicities $\xi_{1,2}$. 
To derive the experimentally accessible cross sections, we must fold in the 
polarized photon fluxes and integrate over the associated energy fractions:
\begin{equation}
\sigma=\int^{x_{max}}~dx_1~\int^{x_{max}}~dx_2~f_\gamma(x_1,\xi_1(x_1),P_{e_1},
P_{\gamma_1})f_\gamma(x_2,\xi_2(x_2),P_{e_2},P_{\gamma_2})
{d\hat \sigma \over {dz}}\,,
\end{equation}
where we explicitly note the dependence of the fluxes on the laser and 
electron beam polarization and the average helicities on the beam energy. We 
remind the reader that the $\xi$'s are also dependent on the initial laser 
and electron beam polarizations. (We also must identify 
$\hat s=s_{e^+e^-}x_1x_2$.)  
In the present case the kinematics require the photon energies to satisfy the 
constraint $\tau=\hat s/s=x_1x_2\geq 4m_S^2/s=\tau_{min}$ which, together with 
the value of $x_{max}$, then determines the lower bounds on both $x_{1,2}$: 
$x_1^{min}=\tau_{min}/x_{max}$ and $x_2^{min}=\tau_{min}/x_1$.

\begin{figure}[htbp]
\centerline{
\psfig{figure=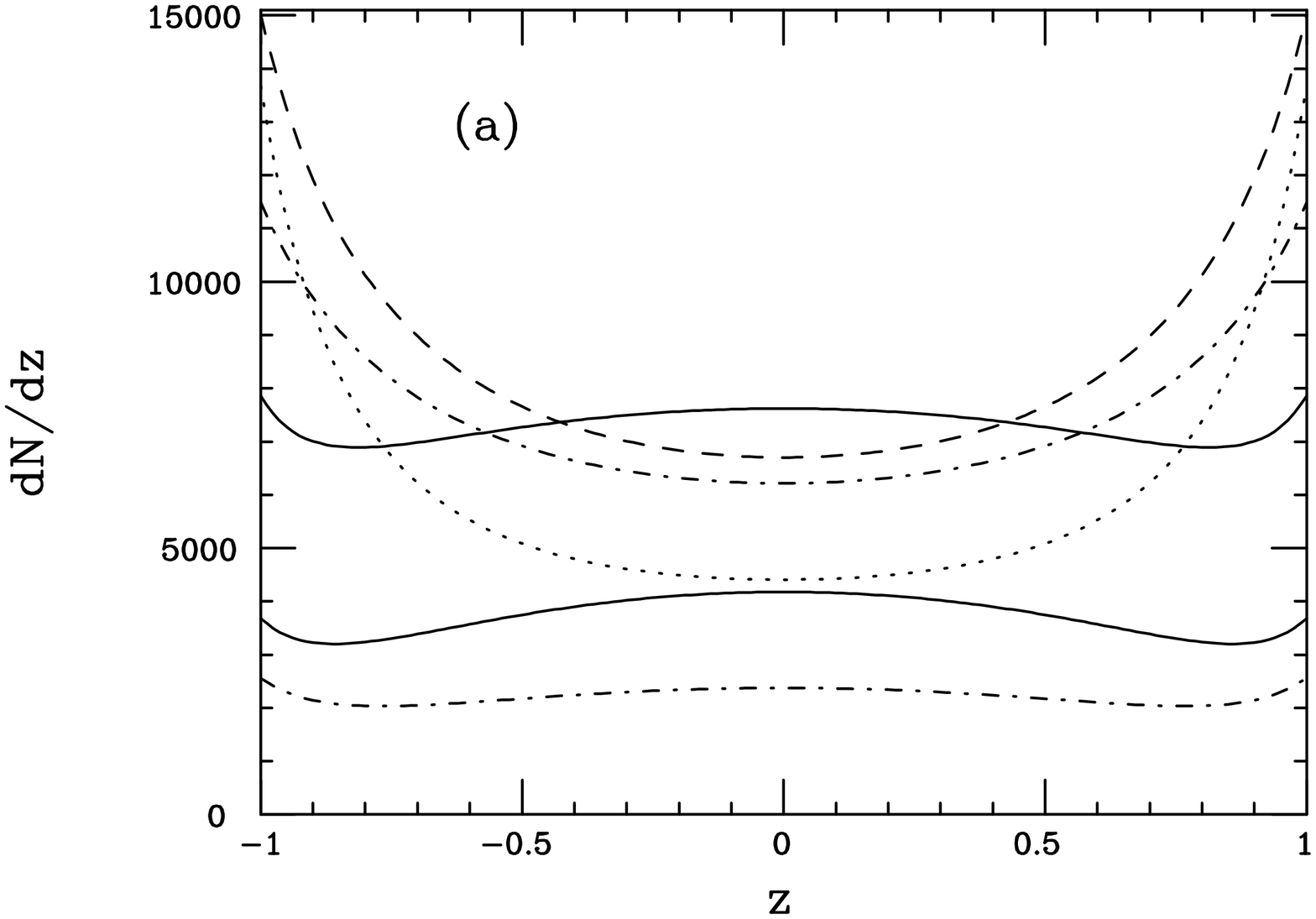,height=9.1cm,width=9.1cm,angle=-90}
\hspace*{-5mm}
\psfig{figure=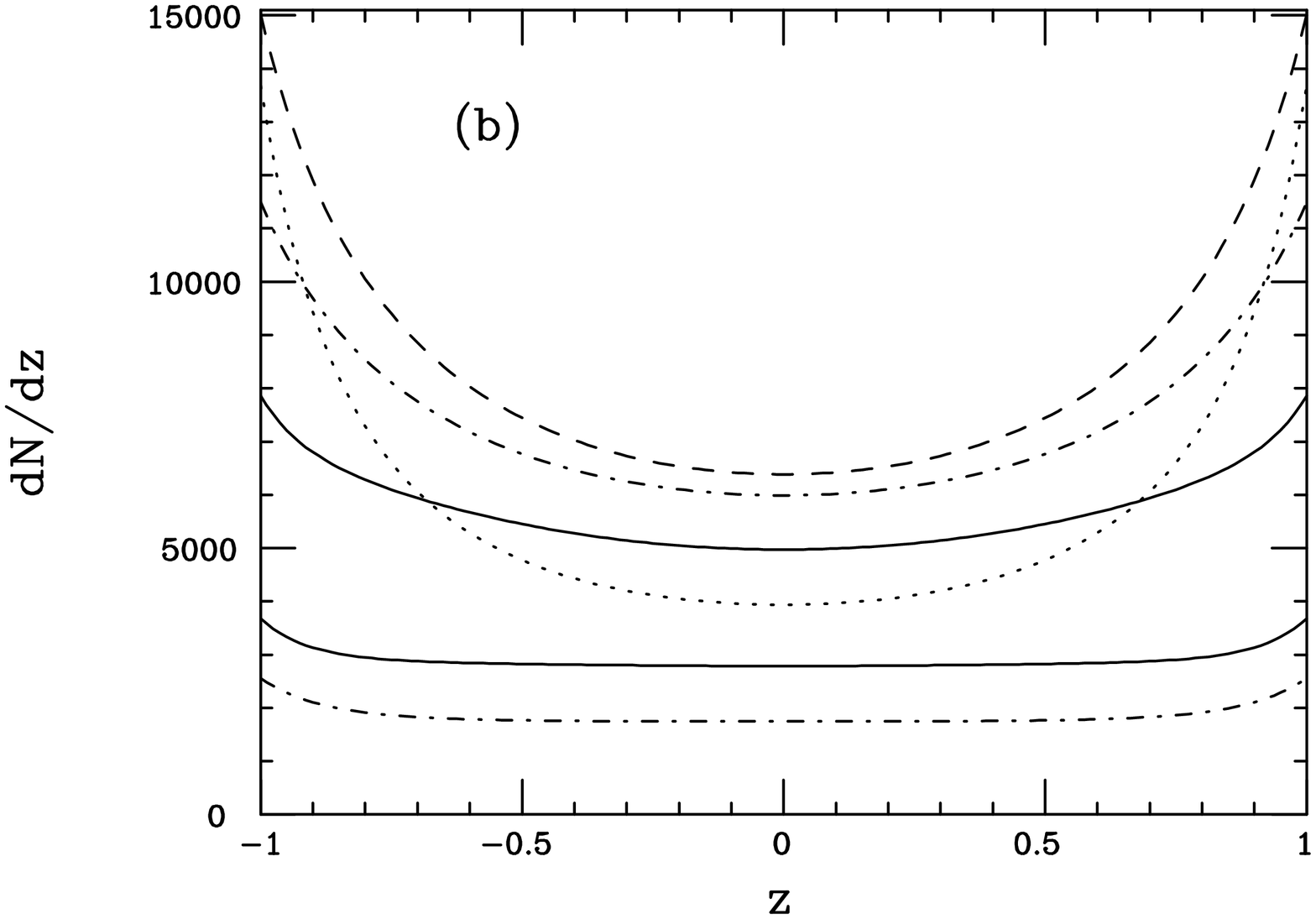,height=9.1cm,width=9.1cm,angle=-90}}
\vspace*{-0.75cm}
\centerline{
\psfig{figure=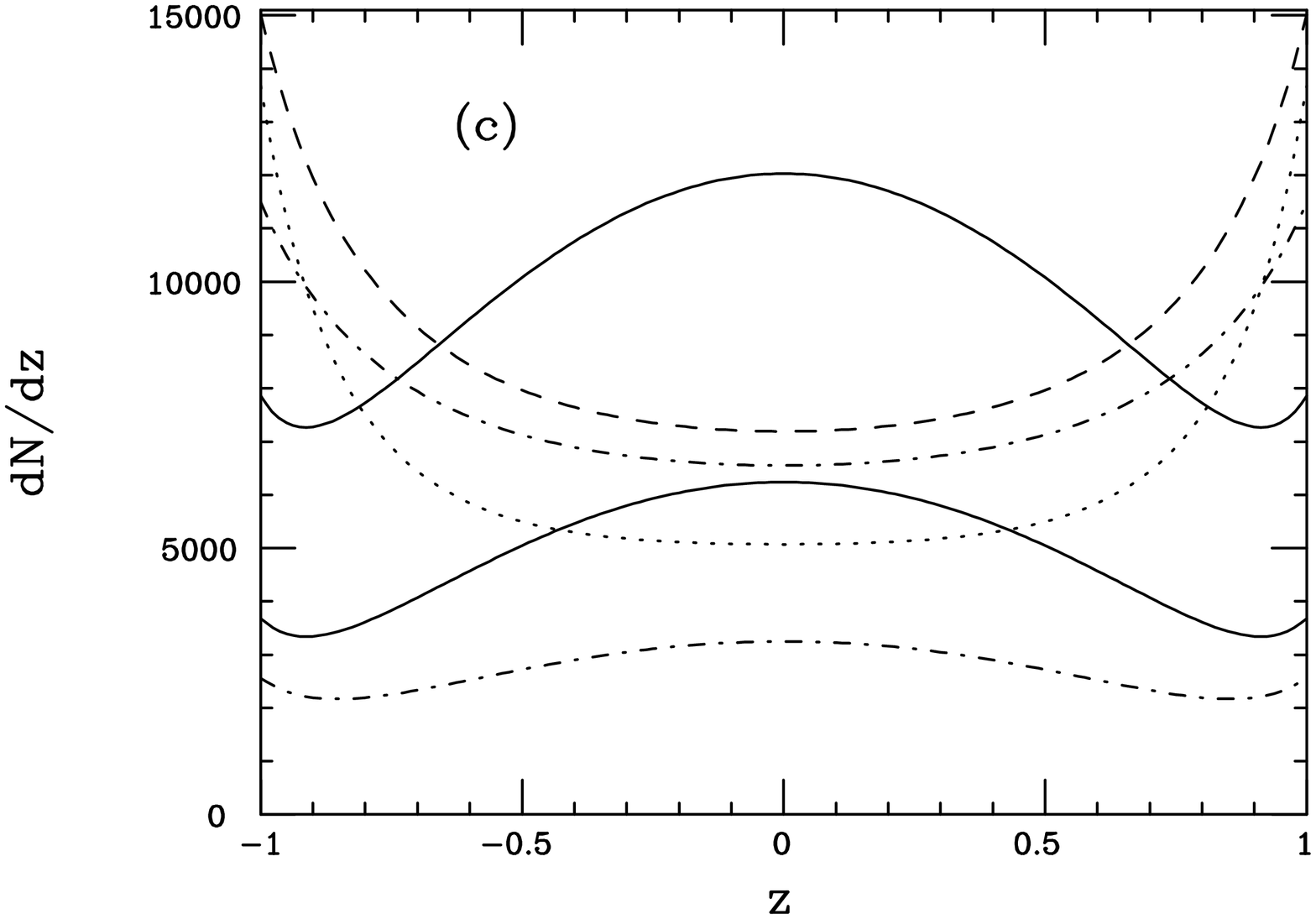,height=9.1cm,width=9.1cm,angle=-90}}
\vspace*{-1cm}
\caption{Un-binned angular distribution as in the previous figure but now 
broken down into the various helicity contributions. A stop mass of 250 GeV 
has been assumed along with an integrated luminosity of 200 $fb^{-1}$ and a 
collider center of mass energy of 1 TeV. (a) is the MSSM while 
(b) and (c) are the results including graviton exchange with $\lambda =\pm 1$ 
with $M_s=2$ TeV. In all panels, the helicities are as follows: $(++++)$ is 
the upper dash-dotted curve, $(+++-)$ is the dashed curve, $(++--)$ is the 
lower dash-dotted curve, $(+-+-)$ is the dotted curve, $(-++-)$ is the upper 
solid curve and $(+---)$ is the lower solid curve.}
\label{fig6}
\end{figure}

We begin as before with the example of stop pair production. We now no 
longer need to specify the $\tilde t_L-\tilde t_R$ mixing angle since the cross 
section depends only on $Q_{\tilde t}$ and $m_{\tilde t}$, though our ability 
to detect and reconstruct this final state may remain dependent on this mixing 
parameter.  For simplicity, we first consider the case of 
{\it unpolarized} photons in which case the resulting binned angular 
distribution is 
shown in Fig.\ref{fig5} for both the standard MSSM scenario and that including 
graviton tower exchange. For the chosen set of parameters it is quite clear 
that new physics is present by the fact that the dip in the angular 
distribution near $90^o$ has been made deeper or more shallow in a 
statistically significant manner. Assuming no deviations from the MSSM 
expectations are observed with this assumed integrated luminosity we estimate 
that the $95\%$ CL lower bound placed on $M_s$ would be $\simeq 3.2$ TeV. As 
we will see below, beam polarization will significantly improve this bound.

What happens when we polarize the colliding 
photons? In this case the results are shown in the 3 panels of Fig.\ref{fig6} 
for the six independent polarization choices. We see immediately that two of 
the cases which yield the largest cross sections, $(++++)$ and $(+++-)$, are 
hardly effected by the contributions due to graviton tower exchange 
whereas those corresponding to both $(-++-)$ and $(+---)$ are quite 
significantly modified. $(++--)$ and $(+-+-)$ are seen to be somewhat less 
affected by K-K exchange but the cross sections are visibly shifted by a 
substantial amount. The reasons why these particular choices are the most 
sensitive to K-K exchange is easily seen by expanding Eq.7 and examining the 
new pieces due to gravitons:
\begin{equation}
\delta {d\hat \sigma\over {dz}} \sim [-4Q^2XY+2Y^2](1-\xi_1\xi_2)\,.
\end{equation}
This means that polarization choices that lead to large negative values of 
the product $\xi_1\xi_2$ in the kinematic region of interest will display the 
greatest 
sensitivity to K-K tower exchange. Indeed both the polarization choices 
$(-++-)$ and $(+---)$ fulfill 
this expectation as can be seen by an explicit calculation. Employing these 
polarizations and assuming no signal for K-K graviton tower exchange is 
observed, we can place a $95\%$ CL bound on $M_s$ assuming an integrated 
luminosity of 200 $fb^{-1}$ of $\simeq 3.8$ TeV--a value 
significantly above that obtained in the unpolarized case which reflects the 
added sensitivity. 

\begin{figure}[htbp]
\centerline{
\psfig{figure=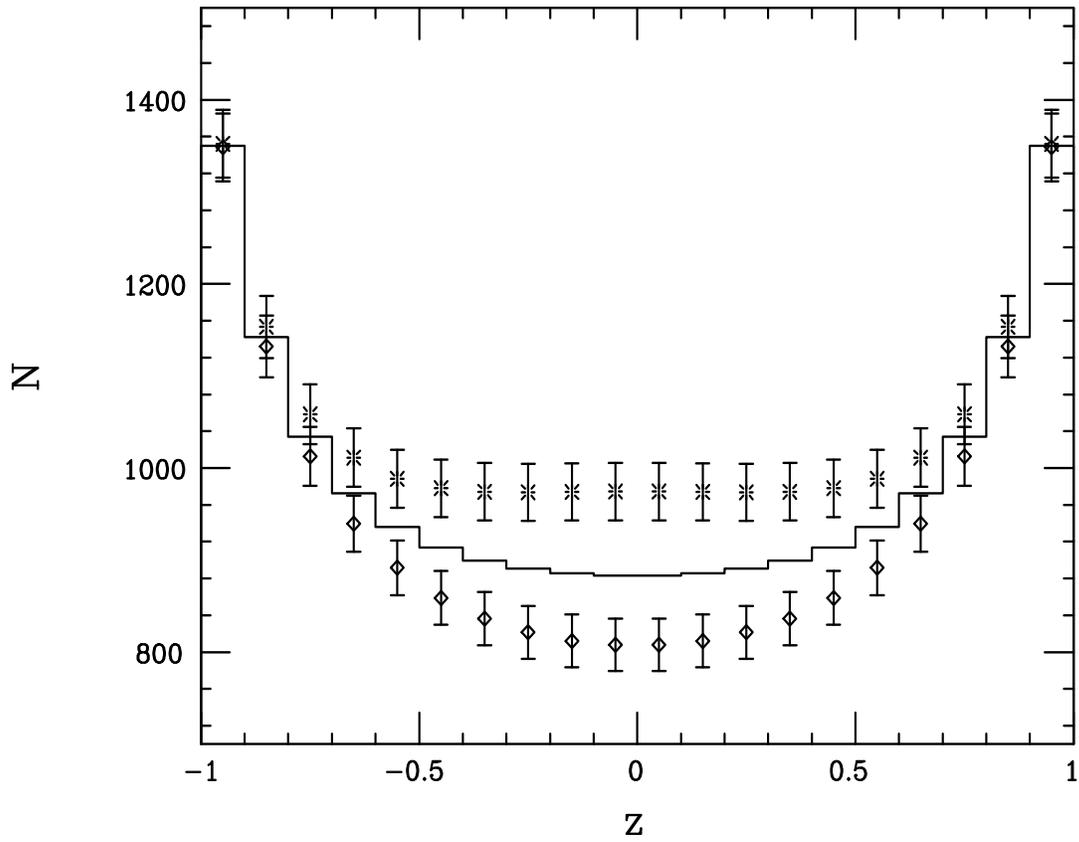,height=14cm,width=17cm,angle=-90}}
\vspace*{-1cm}
\caption[*]{Same as Fig.\ref{fig5} but now for either $H^-$ or $\tilde \ell$ 
pair production by unpolarized photons.}
\label{fig7}
\end{figure}

\begin{figure}[htbp]
\centerline{
\psfig{figure=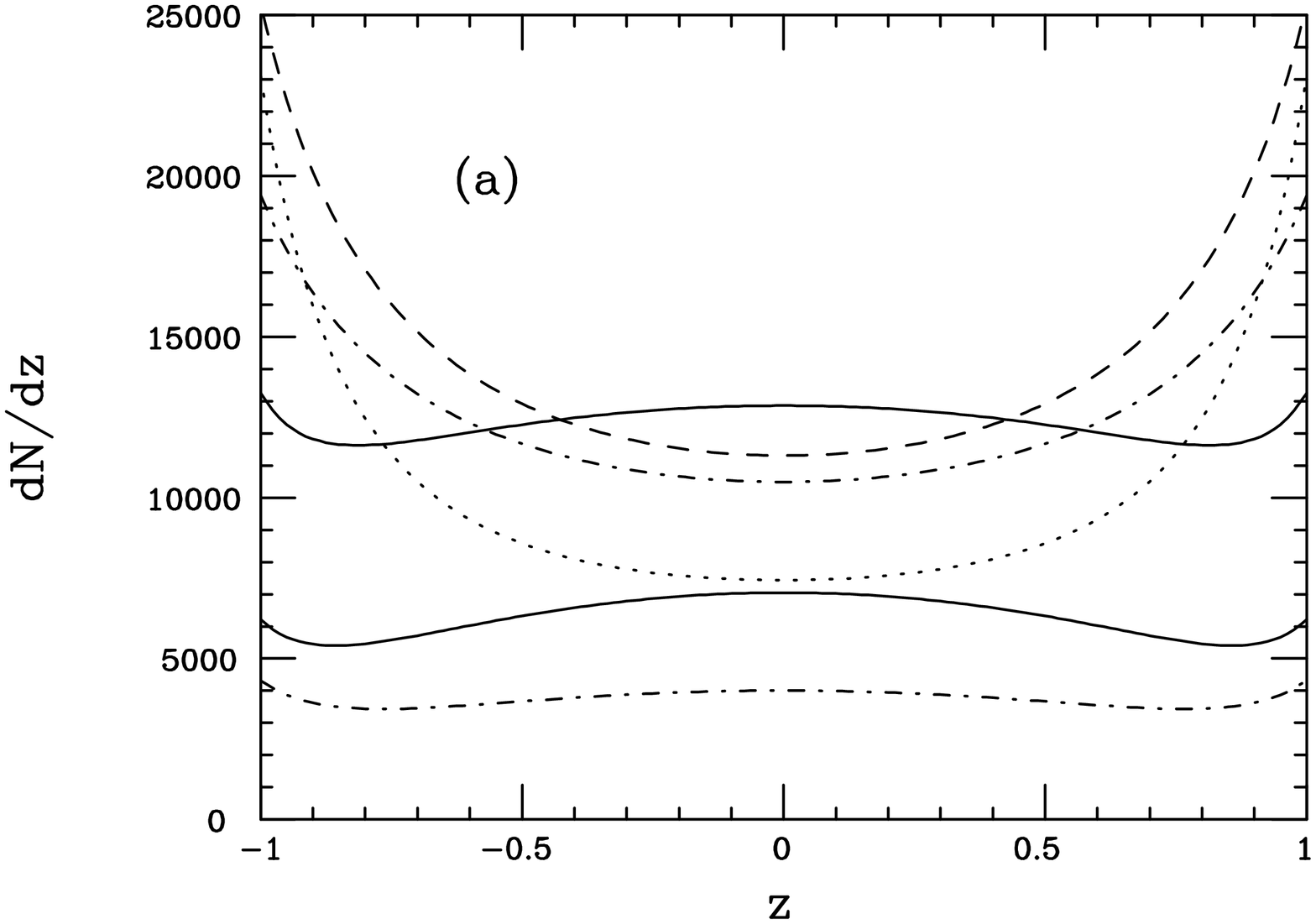,height=9.1cm,width=9.1cm,angle=-90}
\hspace*{-5mm}
\psfig{figure=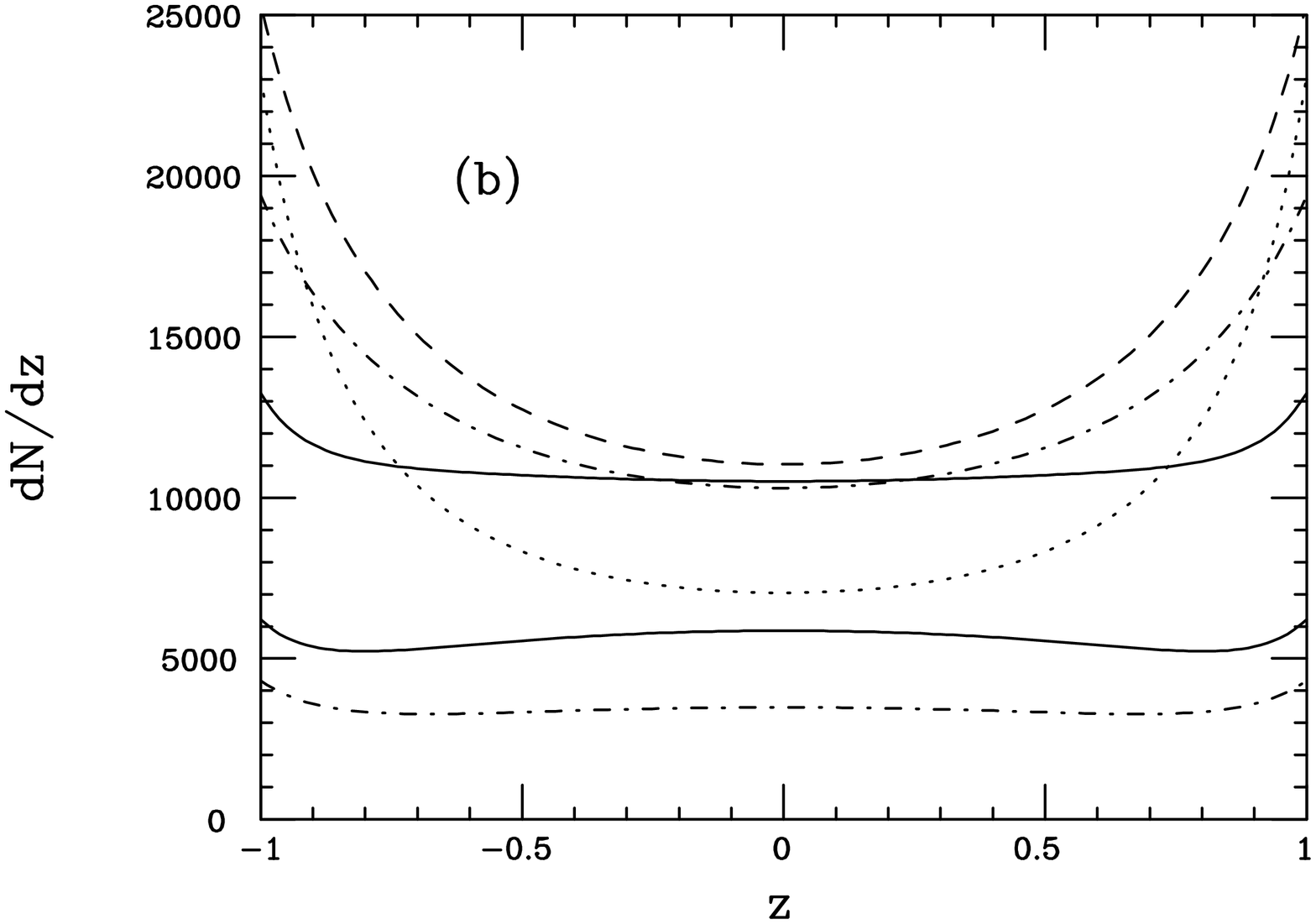,height=9.1cm,width=9.1cm,angle=-90}}
\vspace*{-0.75cm}
\centerline{
\psfig{figure=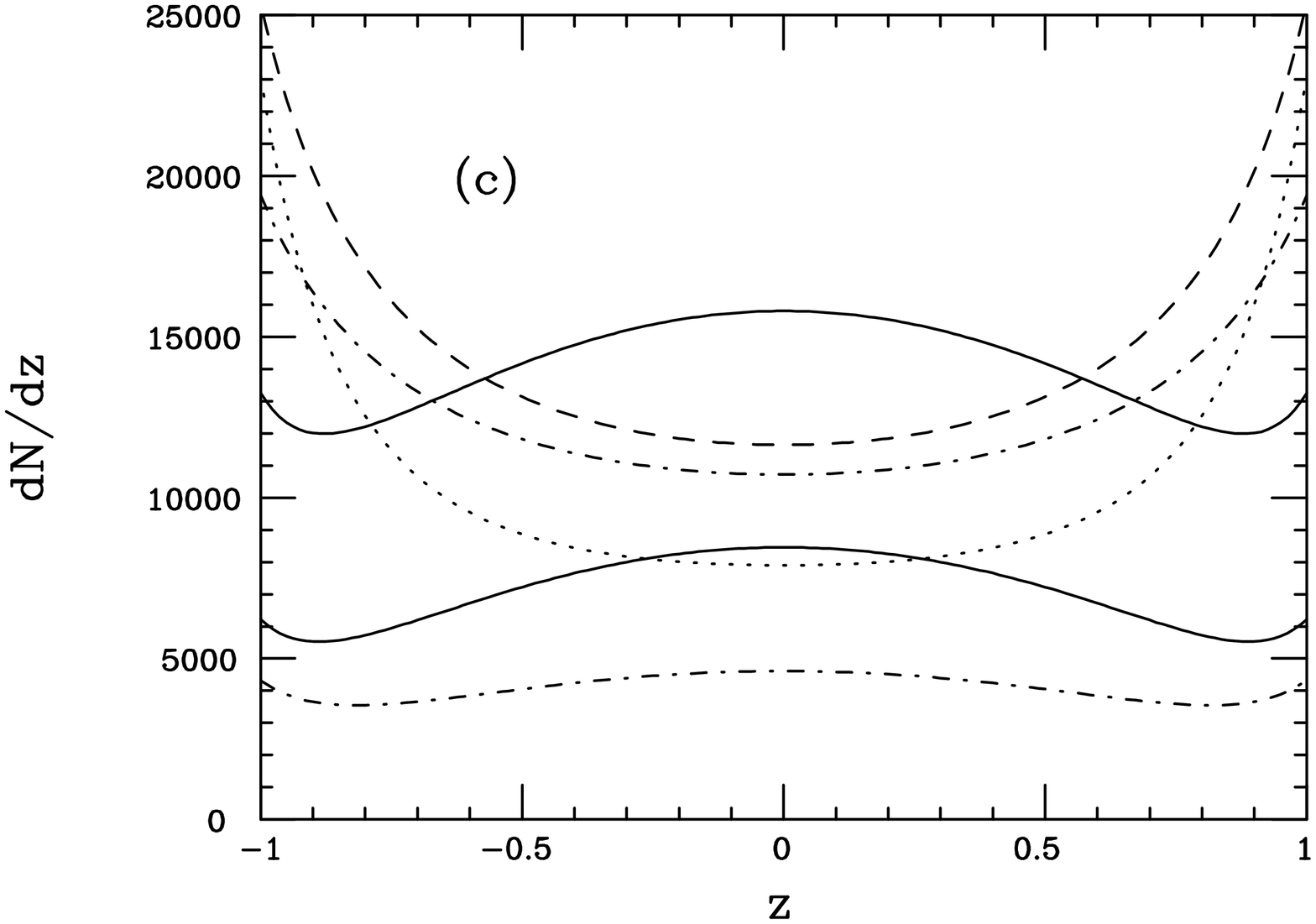,height=9.1cm,width=9.1cm,angle=-90}}
\vspace*{-1cm}
\caption{Same as Fig.\ref{fig6} but now for either $H^-$ or $\tilde \ell$ pair 
production by polarized photons.}
\label{fig8}
\end{figure}

In Figs.~\ref{fig7} and \ref{fig8} we show the corresponding cross sections 
for the production of charged Higgs or slepton pairs using both unpolarized as 
well as polarized photons. These cross sections are all somewhat larger than 
the corresponding ones for stop pairs since, in the exact MSSM limit, the 
total cross section is directly proportional to the product $N_cQ^4$. (Thus 
the results in Fig.\ref{fig7} and Fig.\ref{fig5} for the case of the MSSM 
differ only by an overall constant.) This also implies that 
for fixed $M_s$ the $H^\pm$ and $\tilde \ell$ pair cross sections are overall 
less sensitive to graviton tower exchange due to the larger value of the 
scalar's electric charge, though this is partially compensated 
for by the increase in the available statistics in the case of unpolarized 
beams and nearly completely so when beam polarization is available. (This 
would further imply that $\tilde b$ pair production would be the most 
sensitive to graviton exchange due to the small size of $Q^2=1/9$ in this 
case. Indeed this is true but the reduced size of the MSSM cross section, 
suppressed by $Q^4$, leads to very  small statistical samples.) 
These observations and expectations are indeed 
verified by examining Fig.9. We again see from these figures that the best 
sensitivity to K-K exchanges arises from the particular polarization 
choices $(-++-)$ and $(+---)$. In the case of unpolarized(polarized) beams 
the absence of any deviation from MSSM expectations would lead to a $95\%$ CL 
lower bound on $M_s$ of $\simeq 2.8(3.8)$ TeV for this value of the integrated 
luminosity. 

\vspace*{-0.5cm}
\nn
\vspace*{0.4mm}

\begin{figure}[htbp]
\centerline{
\psfig{figure=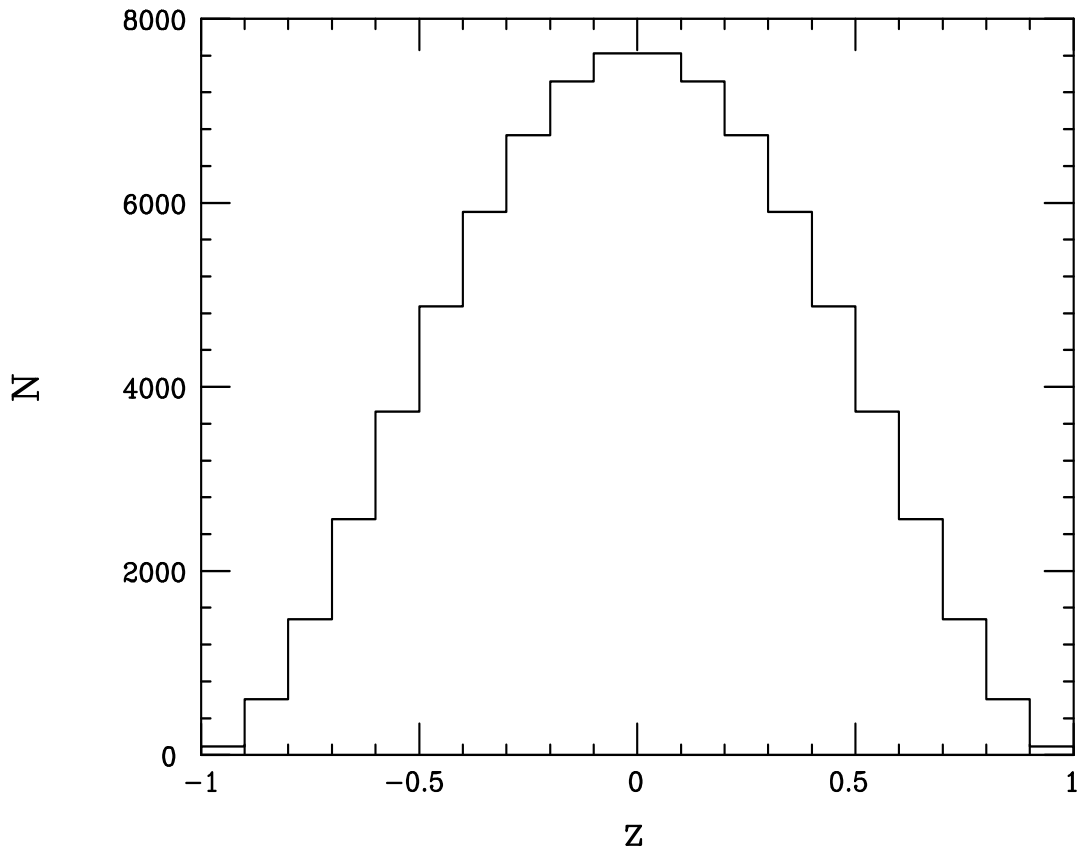,height=14cm,width=17cm,angle=-90}}
\vspace*{-1cm}
\caption[*]{Binned angular distribution for Higgs boson pair production in 
unpolarized $\gamma\gamma$ collisions at a 1 TeV $e^+e^-$ assuming $m_h=130$ 
GeV, $M_s=1$ TeV and an integrated luminosity of 200 $fb^{-1}$.}
\label{fig9}
\end{figure}

As was the case of $e^+e^-$ collisions, the existence of K-K graviton exchange 
now allows for the production of neutral scalars, such as sneutrinos or 
Higgs bosons, in 
$\gamma\gamma$ collisions. The cross section for the production of identical 
Higgs boson pairs is 
trivially obtainable from the above by setting $Q=0$ and dividing by a factor 
of 2 since identical particles now reside in the final state. We obtain
\begin{equation}
{d\hat \sigma \over {dz}}= {\lambda^2K^2\hat s^3\over {32\pi M_s^8}}\beta^5
(1-z^2)^2(1-\xi_1\xi_2)\,,
\end{equation}
following the above notation. As in the case of $e^+e^-$ annihilation, the 
Higgs pairs appearing in the final state for the case of the MSSM can be 
either $hh$, $AA$, or $HH$ but no `off-diagonal' couplings are present. For 
particles of the same mass all of these cross sections are predicted to be 
identical. Such processes cannot occur at the tree level 
in the SM or MSSM but they can appear at one loop{\cite {loop2}} to which we 
will later compare our results. Note that the above expression for the 
differential cross section leads to a very distinctive angular distribution 
$\sim \sin^4 \theta$ for the production of neutral Higgs pairs which only  
populates the amplitude where the two photon helicities are opposite. This 
rather unique angular distribution, resulting from spin-2 exchange, is shown 
for the case of unpolarized photons in Fig.\ref{fig9} for the very simple 
choice $M_s=\sqrt s$=1 TeV and $m_h=130$ GeV. The total integrated cross 
section for these parameters is found to be quite large, 
$\simeq 410$ $fb$ implying the availability of very large rates; results for 
other values of $M_s$ and $\sqrt s$ can be obtained by very simple scalings 
using Eq.11.

\begin{figure}[htbp]
\centerline{
\psfig{figure=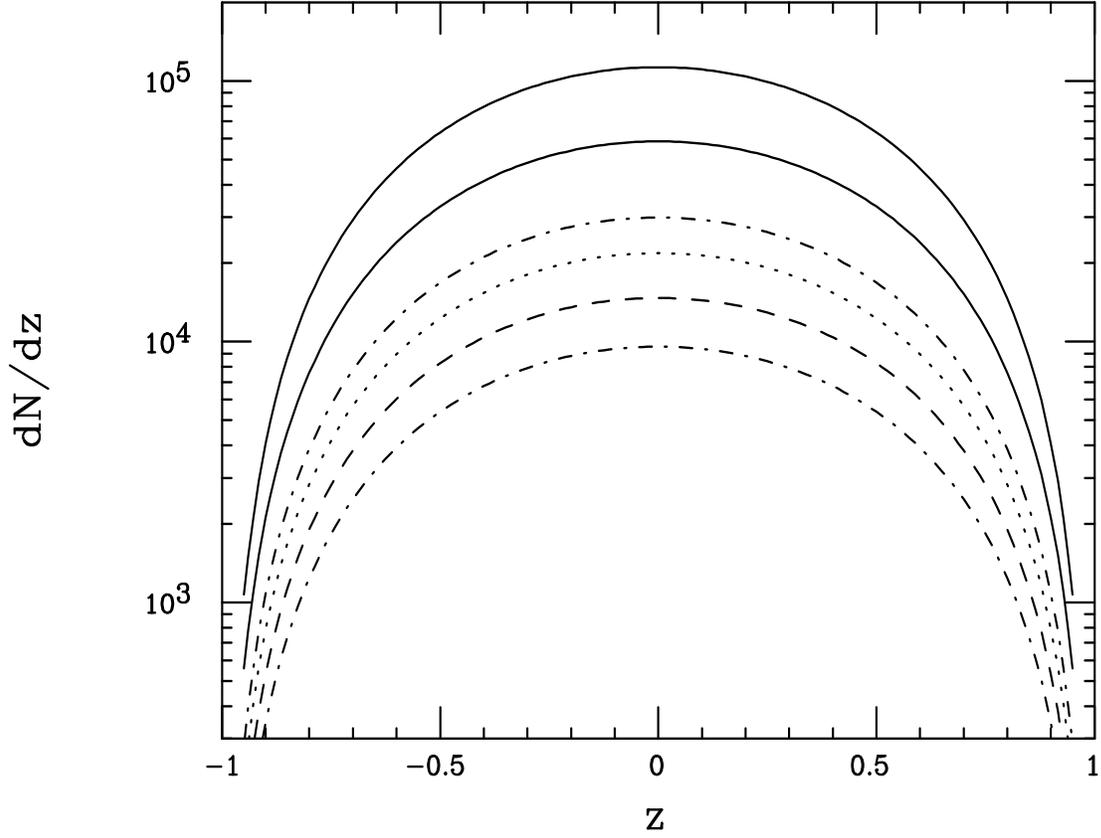,height=14cm,width=17cm,angle=-90}}
\vspace*{-1cm}
\caption[*]{Un-binned angular distribution for the process $\gamma \gamma \to 
2h^0$ using polarized beams for the same parameter choice as in the previous 
figure. From top to bottom the polarization parameter choices are given by 
$(-++-)$, $(+---)$, $(++--)$, $(+-+-)$, $(+++-)$ and $(++++)$, respectively.}
\label{fig10}
\end{figure}

For the case of polarized photons the angular distribution for this process 
is found to be independent of the choice of electron or laser polarizations 
unlike the case of charged scalar pair production and is shown in 
Fig.\ref{fig10}. Here the large rates obtained by the polarization choices 
$(-++-)$ and $(+---)$ are easily explained. The argument essentially parallels 
the explanation for the sensitivities of these polarization choices to K-K 
exchanges. First, both these distributions 
are somewhat larger than the others at values of $\sqrt \tau \geq 0.6$ where 
most of the cross section originates. Second, the dominant enhancement 
arises since both of these polarization 
choices lead to large negative values of the product $\xi_1\xi_2$ in the same 
invariant mass range. From the $1-\xi_1\xi_2$ dependence of the cross section 
this leads to a large enhancement in the production rate. The choice $(++--)$ 
also yields a somewhat large negative value of $\xi_1\xi_2$ and it too is 
seen to be somewhat enhanced.

\vspace*{-0.5cm}
\nn
\vspace*{0.4mm}

How large are these cross sections in comparison to SM and MSSM expectations 
and do their angular distributions and polarization dependencies differ? 
In the SM, the loop induced cross section{\cite {loop2}} obtains contributions 
from both the same and opposite sign photon helicity combinations unlike the 
case of graviton exchange. With $\sqrt s$=1 
TeV and a Higgs mass of 130 GeV a cross section of order 0.2 fb is obtained 
and found to be reasonably insensitive (at the $\pm 50\%$ level) to the choice 
of laser and electron polarizations. In the unpolarized case such a cross 
section is significantly below that obtained from K-K graviton tower exchange 
unless $M_s>3.5$ TeV. As the Higgs boson mass increases the tree-level K-K 
induced rate falls slower than does that induced by SM or MSSM loops. 
The apparent size of this cross section for the 
$CP$-even Higgs in the ordinary
Two-Higgs Doublet extension is quite comparable to that found in the SM 
whereas in the MSSM with large stop mixing the rate can be dramatically 
enhanced{\cite {loop2}}. If stop mixing is reasonably small then the MSSM and 
SM predictions are again comparable.

\vspace*{-0.5cm}
\nn
\vspace*{0.4mm}

\section{Higgs Pair Production at Hadron Colliders}

\begin{figure}[htbp]
\centerline{
\psfig{figure=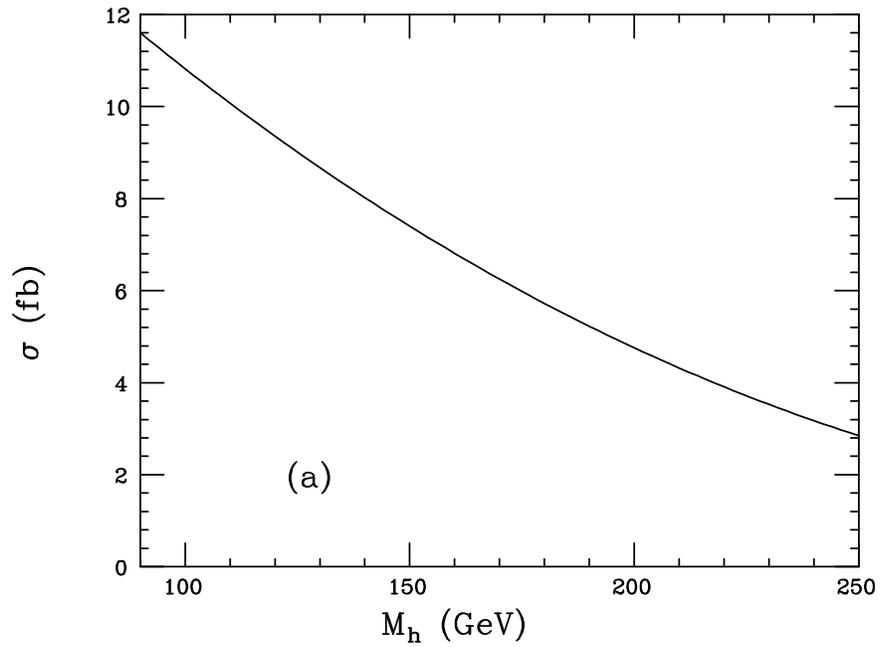,height=10.5cm,width=14cm,angle=-90}}
\vspace*{-10mm}
\centerline{
\psfig{figure=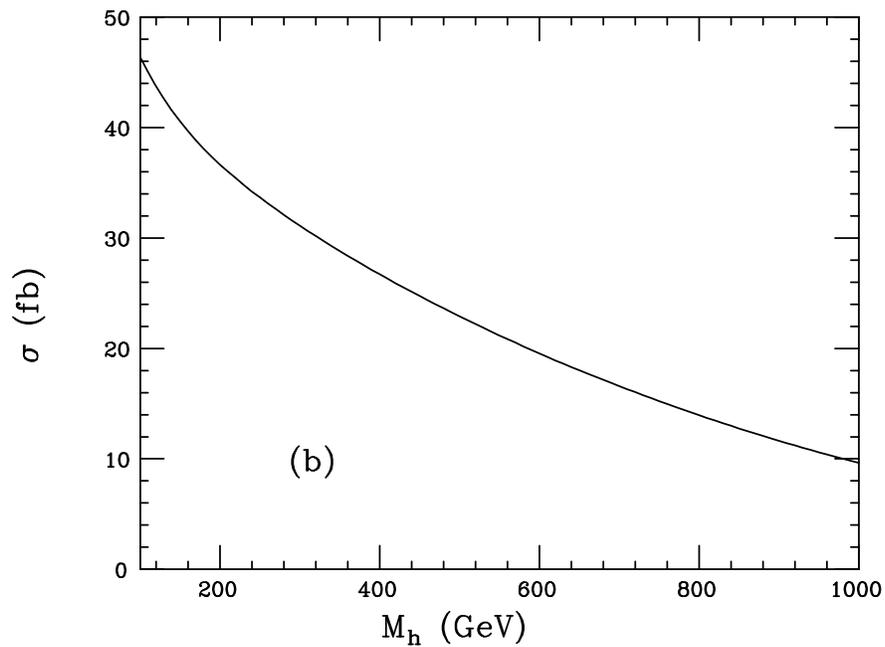,height=10.5cm,width=14cm,angle=-90}}
\vspace*{-0.9cm}
\caption{Leading order production cross sections for Higgs boson pairs at 
(a) TeV II or (b) the LHC as a function of the Higgs boson mass at the tree 
level due to the exchange of a Kaluza-Klein tower of gravitons. We assume 
that $M_s=1(3)$ TeV for the case of the Tevatron(LHC).}
\label{fig11}
\end{figure}

By analogy with the processes $e^+e^-,\gamma\gamma \to 2h^0$ discussed above, 
it will be possible to pair produce Higgs bosons at hadron colliders by very 
similar 
mechanisms: $q\bar q, gg \to 2h^0$. Since the K-K tower of gravitons represents 
a color singlet operator the cross sections for these reactions 
are very easily obtained from the expressions above by dividing by color 
factors of 3 or 8, respectively, and then weighting them with the 
appropriate parton densities. The result of this straightforward analysis 
yields the results shown in Fig.\ref{fig11} at leading order for both the 
Run II Tevatron and the LHC. (We note that NLO QCD corrections may 
significantly increase these rates by as much as a factor of two as they do 
for both the SM and MSSM loop induced processes; we will ignore such effects 
in our discussion below.)
Recall that in either case the cross section scales as 
$\sim M_s^{-8}$ thus falling rapidly as the string scale is increased. Also 
recall that the cross sections for $HH$ or $AA$ production due to graviton 
exchange are numerically identical. 
The final state may, in principle, be observed in either the $4b$ or 
$2b\tau^+\tau^-$ mode. 

The first thing to notice is that at the Tevatron the cross section leads to   
only a very small handful of events for this process (for a representative 
value of $M_s$=1 TeV) even before any cuts are imposed. This implies that if 
$M_s$ is only slightly larger than the assumed value there will be essentially 
no statistics available to discover or observe Higgs boson pairs in 
this channel assuming an integrated luminosity of 2 $fb^{-1}$. This situation 
could change if significantly higher integrated luminosities were to be 
achieved. We note that the current 
lower bound on $M_s$ from the analysis of Hewett{\cite {pheno}} is already in 
excess of 1.0-1.1 TeV or so based on data from LEPII and the Run I of the 
Tevatron. For the LHC, taking 
$M_s=3$ TeV and an integrated luminosity of 100 $fb^{-1}$ a sizeable number 
of events can be obtained over a reasonable range of Higgs boson masses. For 
$M_s=3$ TeV, these cross sections range from somewhat larger to quite 
substantially larger than those obtainable 
in the SM or in the MSSM{\cite {sally}} for $CP$-even scalars, expect in the 
case where the $H\to hh$ channel opens up with the $H$ produced directly 
on-shell. The 
cross section at the LHC for the $CP$-odd pairs in the MSSM is always quite 
small in comparison to the $CP$-even case implying that the K-K contributions 
can be very large and possibly dominant.

It thus appears that if the string scale is not too large then Higgs boson 
pair production via K-K tower graviton exchange will become an exciting 
possibility at the LHC and, if we are very lucky, the Tevatron. 

\vspace*{-0.5cm}
\nn
\vspace*{0.4mm}

\section{Summary and Conclusions}

In this paper we have extended the phenomenological analyses of the ADD 
scenario to a number of new processes involving scalar final states and 
the exchange of a Kaluza-Klein tower of gravitons at various types of 
colliders. The main points of our analysis are as follows: 
\begin{itemize}

\item  As in the case of $e^+e^-\to f\bar f$ processes, the exchange of a K-K 
tower of gravitons in the $s$-channel can significantly modify the angular 
distribution and left-right polarization asymmetry in the case of scalar pair 
production. In particular, graviton exchange leads to a qualitatively flavor 
independent and statistically significant forward-backward asymmetry provided 
the scale $M_s$ is not too far above $\sqrt s$. Such an asymmetry for squarks, 
sleptons and charged Higgs produced in $e^+e^-$ annihilation is essentially 
impossible to mimic by other forms of new physics 
such as a $Z'$, $R$-parity violation or leptoquark exchange. Perhaps more 
exciting, we observed that 
K-K tower exchange leads to the tree-level production of identical 
Higgs boson pairs with a reasonable cross section. In contrast, within the SM 
or MSSM, Higgs pair production in $e^+e^-$ collisions can only occur through 
loops. 

\item  $\gamma \gamma \to S\bar S$ using polarized Compton backscattered 
laser photons offers another window on graviton exchange. As in the case of 
$e^+e^-$ annihilation, K-K towers introduce distortions in the scalar pair 
angular distributions which were shown to be particularly sensitive to the 
polarizations choices made for the lasers and electrons in the initial state. 
Since the SM cross section is here proportional to $Q^4$ the flavor of the 
scalar plays an intricate role in determining the sensitivity to $M_s$: 
scalars with large(small) values of $Q$ have larger SM cross sections hence 
more(less) statistical power. On the otherhand, the {\it fractional} shift in 
the total amplitude due to graviton exchange is clearly smaller(larger) in 
that case. For $\tilde t$ and $\tilde l$ pairs these two effects approximately 
cancel with the help of the additional color factor for $\tilde t$'s.

\item  Graviton tower exchange leads to new operators which can lead to tree 
level processes which cannot occur in either the SM or the MSSM that involve 
scalar pairs in the final state. Here we have considered as particularly 
interesting examples of this phenomena the processes 
$e^+e^-,\gamma\gamma \to 2h^0$ and $\gamma\gamma \to \tilde \nu \tilde \nu^*$ 
at lepton colliders and $q\bar q,gg \to 2h^0$ 
at hadron colliders. These reactions were shown to have potentially 
significant cross sections and are likely to be easily separable from the 
corresponding loop induced processes in the SM and MSSM. 

\item  Scalar pair production at colliders, though not the likely channel for 
the discovery of new operators associated with extra dimensions, provides an 
additional channel with which to explore the implications of theories of low 
scale quantum gravity.

\end{itemize}
New dimensions may soon make their presence known at existing and/or future 
colliders. Such a discovery would revolutionize the way we think of physics 
beyond the electroweak scale.

\noindent{\Large\bf Acknowledgements}

The author would like to thank J.L. Hewett, N. Arkani-Hamed, J. Wells, T. Han, 
J. Lykken, M. Schmaltz and H. Davoudiasl for extra-dimensional and off the 
wall discussions related to this work.

\newpage

%
\def\MPL #1 #2 #3 {Mod. Phys. Lett. {\bf#1},\ #2 (#3)}
\def\NPB #1 #2 #3 {Nucl. Phys. {\bf#1},\ #2 (#3)}
\def\PLB #1 #2 #3 {Phys. Lett. {\bf#1},\ #2 (#3)}
\def\PR #1 #2 #3 {Phys. Rep. {\bf#1},\ #2 (#3)}
\def\PRD #1 #2 #3 {Phys. Rev. {\bf#1},\ #2 (#3)}
\def\PRL #1 #2 #3 {Phys. Rev. Lett. {\bf#1},\ #2 (#3)}
\def\RMP #1 #2 #3 {Rev. Mod. Phys. {\bf#1},\ #2 (#3)}
\def\NIM #1 #2 #3 {Nuc. Inst. Meth. {\bf#1},\ #2 (#3)}
\def\ZPC #1 #2 #3 {Z. Phys. {\bf#1},\ #2 (#3)}
\def\EJPC #1 #2 #3 {E. Phys. J. {\bf#1},\ #2 (#3)}
\def\IJMP #1 #2 #3 {Int. J. Mod. Phys. {\bf#1},\ #2 (#3)}


\begin{thebibliography}{99}

\bibitem{nima}
N. Arkani-Hamed, S. Dimopoulos and G. Dvali, \PLB B429 263 1998 ~and 
hep-ph/9807344; I. Antoniadis, N. Arkani-Hamed, S. Dimopoulos and G. Dvali, 
\PLB B436 257 1998; N. Arkani-Hamed, S. Dimopoulos and J. March-Russell, 
hep-th/9809124; P.C. Argyres, S. Dimopoulos and J. March-Russell, 
hep-th/9808138; Z. Berezhiani and G. Dvali, hep-ph/9811378; N. Arkani-Hamed 
and S. Dimopoulos, hep-ph/9811353; Z. Kakushadze, hep-th/9811193 and 
hep-th/9812163; N. Arkani-Hamed \etal, hep-ph/9811448; 
G. Dvali and S.-H.H. Tye, hep-ph/9812483. See also, 
G. Shiu and S.-H. H. Tye, \PRD D58 106007 1998 ;
Z. Kakushadze and S.-H. H. Tye, hep-th/9809147;
I. Antoniadis, \PLB B246 377 1990 ;
J. Lykken, \PRD D54 3693 1996 ;
E. Witten, \NPB B471 135 1996 ;
P. Horava and E. Witten, \NPB B460 506 1996 ~and \NPB B475 94 1996 ;
K.R. Dienes, E. Dudas and T. Gherghetta, \PLB B436 55 1998 ,~hep-ph/9803466, 
hep-ph/9806292 and hep-ph/9807522.
%
\bibitem{test}
J.C. Long, H.W. Chan and J.C. Price, hep-ph/9805217.
%
\bibitem{astro}
S. Cullen and M. Perelstein, hep-ph/9903422 and references therein. 
%
\bibitem{pheno}
G.F. Guidice, R. Rattazzi and J.D. Wells, hep-ph/9811291;
T. Han, J.D. Lykken and R.J. Zhang, hep-ph/9811350;
J.L. Hewett, hep-ph/9811356;
E.A. Mirabelli, M. Perelstein and M.E. Peskin, hep-ph/9811337;
P. Mathews, S. Raychaudhuri and K. Sridhar, hep-ph/9811501 and hep-ph/9812486; 
S. Nussinov and R.E. Shrock, hep-ph/9811323;
T.G. Rizzo, hep-ph/9901209 and hep-ph/9902273; 
K. Agashe and N.G. Deshpande, hep-ph/9902263; 
M.L. Graesser, hep-ph/9902310;
K.Cheung and W.-Y. Keung, hep-ph/9903294;
N. Arkani-Hamed and M. Schmaltz, hep-ph/9903417.
%
\bibitem{loop1}
See, for example, K.J.F. Gaemers and F. Hoogeveen, \ZPC C26 249 1984 ;
A. Djouadi, V. Driesen and C. J\"unger, \PRD D54 759 1996 .
%
\bibitem{telnov}
For a recent review of $\gamma \gamma$ colliders, photon distributions and 
original references, see V. Telnov, hep-ex/9810019. For more details, see 
I.F. Ginzburg \etal, \NIM 205 47 1983 , \NIM 219 5 1984 , 
\NIM A294 2 1990 ~and \NIM A355 3 1995 ;
V.I. Telnov, \NIM A294 72 1990 ;
D.L. Bordon, D.A. Bauer and D.O. Caldwell, SLAC-PUB-5715 (1992). 
%
\bibitem{chak}
Our helicity sign convention is that used by S. Chakrabarti \etal, 
\PLB B434 347 1998 .
%
\bibitem{loop2}
For a discussion of Higgs pair production in the SM and MSSM at the one loop 
level see, G.V. Jikia and Yu.F. Pirogov, \PLB B283 135 1992 ;
G.V. Jikia, \NPB B412 57 1994 ;
S.La-Zhen and L. Yao-Yang, \PRD D54 3563 1996 ;
S.H. Zhu, C.S. Li and C.S. Gao, \PRD D58 015006 1998 .
%
\bibitem{sally}
For a very recent analysis, see S.Dawson, S. Dittmaier and M. Spira, 
\PRD D58 115012 1998 ~and references therein. 
%
\end{thebibliography}
\end{document}